\documentclass[journal]{IEEEtran}
\newcommand{\ee}{\epsilon}		
\newcommand{\E}{\mathrm{E}}

\newcommand{\Var}{\mathrm{Var}}

\usepackage{epsf,psfrag,amssymb,amsfonts,color,cite,fancybox}
\usepackage[mathscr]{eucal}
\usepackage[dvips]{graphicx}
\usepackage{caption}
\usepackage{subcaption}
\usepackage{ifpdf}
\usepackage{longtable}	
\usepackage{multicol}
\usepackage{float}
\usepackage{epstopdf}
\usepackage{bm}

\ifCLASSINFOpdf

\else

\fi

\usepackage[cmex10]{amsmath}
\usepackage[caption=false,font=footnotesize]{subfig}

\begin{document}

\title{Data-Driven Diagnostics of Mechanism and Source of Sustained Oscillations}

\author{Xiaozhe Wang,~\IEEEmembership{Member,~IEEE,}
        Konstantin Turitsyn,~\IEEEmembership{Member,~IEEE.}
\thanks{This work was supported by the MIT/Skoltech initiative and the Ministry of Education and Science of Russian Federation, Grant agreement No. 14.615.21.0001, Grant identification code: RFMEFI61514X0001.}
\thanks{Xiaozhe Wang and Konstantin Turitsyn are with the Department of Mechanical Engineering, MIT, Cambridge, MA, 02139. email: xw264@cornell.edu, turitsyn@mit.edu.}
}

\maketitle

\begin{abstract}
Sustained oscillations observed in power systems can damage equipment, degrade the power quality and increase the risks of cascading blackouts. There are several mechanisms that can give rise to oscillations, each requiring different countermeasure to suppress or eliminate the oscillation. This work develops mathematical framework for analysis of sustained oscillations and identifies statistical signatures of each mechanism, based on which a novel oscillation diagnosis method is developed via real-time processing of phasor measurement units (PMUs) data. Case studies show that the proposed method can accurately identify the exact mechanism for sustained oscillation, and meanwhile provide insightful information to locate the oscillation sources.
\end{abstract}


\IEEEpeerreviewmaketitle

\section{Introduction}
\IEEEPARstart{S}USTAINED low frequency oscillations are one of major concerns to power system operation. Oscillations cause problems for power quality and can potentially damage power grid equipment or activate protective equipment. In most severe scenarios, growing oscillations may lead to catastrophic blackouts \cite{Taylor:1999}.

Generally speaking, there are three mechanisms of power system oscillations. Firstly, the oscillation can appear due to weak damping arising from high-gain fast exciters, long transmission lines, or high transmission powers \cite{Chen:2013}-\cite{Kundur:book}. This kind of oscillation can be quenched by appropriate power system stabilizer (PSS) tuning, intertie line controls, and generator power reduction. Second mechanism of oscillation appearance is attributed to supercritical Hopf bifurcation such that a stable limit cycle is born \cite{Ajjarapu:1992}-\cite{Liaw:2005}. Unlike the weakly damped oscillation which can be analyzed by linearizing the system, the limit cycle is an essentially nonlinear behavior which only exists in nonlinear systems.
The emergence of limit cycle implies that the stability property of the system has changed, and in order to push the system back to stability, operation point may need to be reset. In addition, another mechanism of oscillation is forced oscillation, which is excited by external periodic disturbance including cyclic loads, control loops in power plants, diesel generators, etc. \cite{Magdy:1990}-\cite{Vournas:1991}. The most effective countermeasure is to locate and separate the external disturbance from the system. In summary, there are different mechanisms of power system oscillations, and different countermeasures need to be adopted accordingly. Hence, detecting power system oscillations and diagnosing the corresponding mechanism are of utmost importance in power system monitoring, and of great significance to ensure the secure operation of any power system.

Phasor measurement units (PMUs) have been widely deployed in power grids to provide system states and dynamics in real time \cite{Chen:2014}\cite{Bialek:2008}. The high quality PMU data provides invaluable information to enable oscillation diagnosis. Regarding the weakly damped oscillation, different methods to identify oscillation modes and damping ratios are proposed including Prony analysis \cite{Scharf:1990}, frequency domain decomposition analysis \cite{Venkatasubramanian:2008}, subspace identification method \cite{Venkatasubramanian:2014}, 
robust RLS methods \cite{Zhou:2007} and so forth. Hopf bifurcation has been studied in terms of dynamic stability. Hopf bifurcation not only involves with the oscillation, but also closely relates to the voltage stability. The previous studies show that voltage collapse may arise from the existence of Hopf bifurcation, which is prior to the appearance of saddle-node bifurcation \cite{Ajjarapu:1992}-\cite{Liaw:2005}. In addition, forced oscillation has also been studied by exploring frequency domain techniques \cite{Magdy:1990}-\cite{Vournas:1991}. It has been shown that if the forced oscillation is close to the natural frequencies, resonance may be observed which leads to severe consequences \cite{Dong:2010}\cite{Rostamkolai:1994}\cite{Vournas:1991}. Even though each type of the oscillation has been widely investigated, little effort has been made to distinguish the three mechanisms from time series data such that the right control actions can be executed. The challenges of oscillation diagnosis include low signal-to-noise ratio (SNR) especially when the amplitude of oscillation is small while load fluctuation and noise intensity are large, as well as similar characteristics of time series data. Specifically, the time series data of oscillations with different mechanisms all looks alike to each other.

This paper proposes a novel method to diagnose the mechanisms of sustained oscillations using PMU data. A unified mathematical framework to describe sustained oscillation is developed, under which the statistical signatures of different models are explored. It is shown that sustained oscillations of different mechanisms exhibit distinct statistical signatures including the kurtosis and the power spectral density, based on which an oscillation diagnosis method is developed. Numerical examples show that the proposed diagnosis method is able to accurately identify the exact mechanism for oscillation even when the SNR is low, and meanwhile provides insightful information to locate the oscillation sources, which are crucial to implement the right control actions in a timely manner. Note that the goals and focus of the proposed method and conventional waveform analysis techniques are different and complementary to each other. The proposed method is to diagnose different oscillation mechanisms and locate the potential sources, whereas the conventional waveform analysis techniques focus on analyzing damping, frequency, oscillation mode shapes, etc. By nature, the traditional waveform analysis does not distinguish between different oscillation mechanisms.

The paper is organized as follows. Section \ref{sectionmodel} describes a unified framework under which the mathematical model for each oscillation mechanism is developed. Section \ref{sectionstatistical signatures} presents the statistical signatures for each oscillation mechanism based on which an oscillation diagnosis method is proposed. Case studies are presented in Section \ref{sectioncasestudies} to demonstrate the accuracy and feasibility of the proposed method. Conclusions and perspectives are given in Section \ref{sectionconclusion}.

\section{mechanisms of sustained oscillations}\label{sectionmodel}
The power system dynamic model can be described as:
\begin{eqnarray}
\dot{\bm{x}}&=&\bm{f}(\bm{x,y})\label{fast ode}\\
\bm{0}&=&\bm{g}(\bm{x,y,u})\label{algebraic eqn}
\end{eqnarray}
Equation (\ref{fast ode}) describes dynamics of generators and their associated control as well as load dynamics, and (\ref{algebraic eqn}) describes the electrical transmission system and the internal static behaviors of passive devices. $\bm{f}$ and $\bm{g}$ are continuous functions; vectors $\bm{x}\in\mathbb{R}^{n_{\bm{x}}}$ and $\bm{y}\in\mathbb{R}^{n_{\bm{y}}}$ are the corresponding state variables (generator rotor angles, rotor speeds, etc) and algebraic variables (bus voltages, bus angles, etc), and $\bm{u}$ is the vector describing stochastic behaviors in real-world power systems. The stochastic perturbations can be originated by loads variations, renewable energy power injections, transient rotor vibrations of synchronous machines, measurement errors of control devices, etc. \cite{Milano:2013}\cite{Hines:2015}. In this paper, we are interested in the stochastic perturbations like load variations and renewable energy power injections, which can be modeled as the Ornstein-Uhlenbeck process, which is stationary, Gaussian and Markovian \cite{Hines:2015}\cite{Soder:2010}:
\begin{equation}
\dot{\bm{u}}=-C\bm{u}+{{\sigma}}\bm{\bar{\xi}}\label{sto eqn}
\end{equation}
where $C=\mbox{diag}\{\alpha_1, \alpha_2,\dots,\alpha_{n_u}\}$ which relates to the correlation times of the stochastic processes \cite{Hines:2015}, $\bm{\bar{\xi}}$ is a vector of independent standard Gaussian random variables , $\sigma^2$ is the intensity of noise.

Linearizing (\ref{fast ode}) and (\ref{sto eqn}), and also eliminating algebraic variables $\bm{y}$ from (\ref{algebraic eqn}), the stochastic power system model takes the form:
\begin{eqnarray}
\dot{\bm{v}}=A\bm{v}+{\sigma} B\bm{\bar{\xi}} \label{compactpowersystem}
\end{eqnarray}
where $\bm{v}=[\delta \bm{x},\delta \bm{u}]^T$, $A=\left[\begin{array}{cc}{f_x-f_yg_y^{-1}g_x}&{-f_yg_y^{-1}g_u}\\0&-C\end{array}\right]$, $B=[0,I_{n_{\bm{u}}}]^T$.

In the rest of the paper, we focus on the stochastic model described in (\ref{compactpowersystem}), where $\bm{v}$ is a vector Ornstein-Uhlenbeck process.

As discussed in the introduction, sustained oscillations of power systems can be produced by several vastly different mechanisms. In the sections below we discuss possible origins of oscillations and develop the corresponding mathematical models.

\subsection{Weakly Damped Oscillation}
Weakly damped oscillation can be classified into local mode oscillations, and interarea mode oscillations. Local mode oscillations are usually caused by automatic voltage regulators (AVRs) operating at high output and feeding into weak transmission networks; Interarea mode are associated with weak transmission links and heavy power transfers. Power system stabilizers (PSSs) are the most common means of enhancing the damping and suppressing the oscillations.

Mathematically, weak damping means that all eigenvalues of $A$ in (\ref{compactpowersystem}) are still in the left half plane, whereas the principal eigenvalue which has the minimum absolute value has been very close to the imaginary axis. According to normal form analysis \cite{Vittal:1997}, the dynamics of system (\ref{compactpowersystem}) can be decomposed into dynamics of individual modes, and we focus on the dynamics of the least stable mode described as below:
\begin{eqnarray}
\dot{z}=(-\gamma+i\omega_0)z+\sigma\xi\label{WD}
\end{eqnarray}
where $\gamma>0$ and $\gamma$ is close to zero.

Let $z=x+iy$, $\xi=\xi_x+i\xi_y$, and represent (\ref{WD}) in Cartesian coordinate, we have:
\begin{equation}
\left(\begin{array}{c}\dot{x}\\\dot{y}\end{array}\right)=\left(\begin{array}{cc}-\gamma&-\omega_0\\ \omega_0&-\gamma\end{array}\right)\left(\begin{array}{c}{x}\\{y}\end{array}\right)+\sigma\left(\begin{array}{c}\xi_x\\ \xi_y\end{array}\right)\label{WDxy}
\end{equation}

The processes $x(t)$, $y(t)$ and $z(t)$ are Ornstein-Uhlenbeck process which is stationary, Gaussian and Markovian. Fig \ref{timeseriesWD} shows a typical sample path of $x$ in system (\ref{WDxy}), where $\gamma=0.02$, $\omega_0=0.3\pi$, $\sigma=0.01$.

\subsection{Limit Cycle}
Supercritical Hopf bifurcation, which leads to the emergence of a stable limit cycle, is another mechanism of power system oscillation. Unlike the weakly damped oscillation, the occurrence of Hopf bifurcation indicates that the equilibrium point of system has already lost its stability. The oscillation due to Hopf bifurcation can be regarded as an early warning of voltage collapse, since Hopf bifurcation usually occurs before saddle node bifurcation which results in the final voltage collapse \cite{Ajjarapu:1992}. Countermeasures like power reschedule, load tap changer blocking, and other emergency control may be needed to stop the voltage degradation.

Limit cycles are inherently nonlinear phenomena that can be described only via nonlinear equations. Near the Hopf bifurcation point, the normal form of the stochastic system can be generally represented as:
\begin{equation}
\dot z=(\gamma+i\omega_h)z-|z|^2z+\sigma\xi\label{normal form}
\end{equation}
Here both $z=x+iy$ coincides with the amplitude of the most unstable mode in the leading order and $\xi=\xi_x+i\xi_y$ is the same noise term as in linear case. It is possible to represent (\ref{normal form}) in Cartesian coordinates:
\begin{eqnarray}
\left(\begin{array}{l}\dot{x}\\\dot{y}\end{array}\right)&=&\left(\begin{array}{cc}\gamma&-\omega_h\nonumber\\ \omega_h&\gamma\end{array}\right)\left(\begin{array}{c}{x}\\{y}\end{array}\right)\\
&&-(x^2+y^2)\left(\begin{array}{c}{x}\\{y}\end{array}\right)+\sigma\left(\begin{array}{c}\xi_x\\ \xi_y\end{array}\right)\label{LCxy}
\end{eqnarray}
Fig. \ref{timeseriesLC} shows a typical sample path of $x$ in system (\ref{LCxy}), where $\gamma=0.01$, $\omega_h=0.3\pi$, and $\sigma=0.01$.

\subsection{Forced Oscillation}
Another type of power system oscillation is raised by forced oscillation from cyclic loads, control loops in power plants, diesel or hydro generators, wind turbines, etc. \cite{Vanfretti:2012}-\cite{Vournas:1991}. This kind of oscillation is not a result of the general dynamics of power system, instead, it is caused by an external forcing with a distinct oscillatory behavior. Particularly, forced oscillation may lead to voltage flickering when the frequency is around 10Hz where the human eye is most sensitive; forced oscillation near the natural modes may result in resonance, and small disturbance is then amplified and expanded rapidly in the whole power system. The most effective countermeasure against forced oscillation is to locate and remove the external oscillation source.

Mathematically, the system undergoing forced oscillation can be described as:
\begin{eqnarray}
\dot{z}=(-\gamma+i\omega_0)z+Fe^{i\Omega t}+\sigma\xi\label{forced}
\end{eqnarray}
where $\omega_0$ is the natural frequency and $\Omega$ is the forced frequency.

\begin{figure}[!ht]
\centering
\begin{subfigure}[t]{0.5\linewidth}
\includegraphics[width=1.8in ,keepaspectratio=true,angle=0]{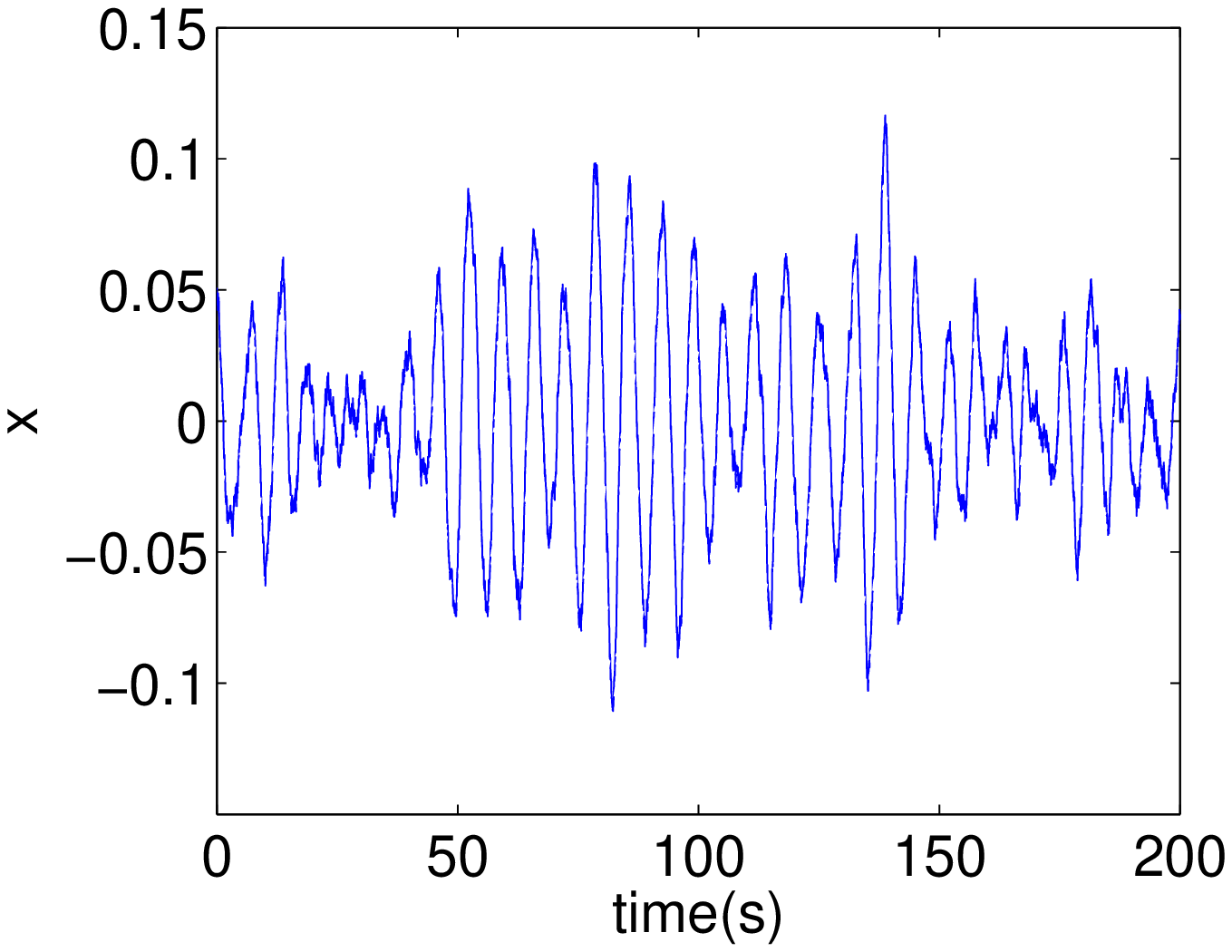}
\caption{}\label{timeseriesWD}
\end{subfigure}%
\begin{subfigure}[t]{0.5\linewidth}
\includegraphics[width=1.8in ,keepaspectratio=true,angle=0]{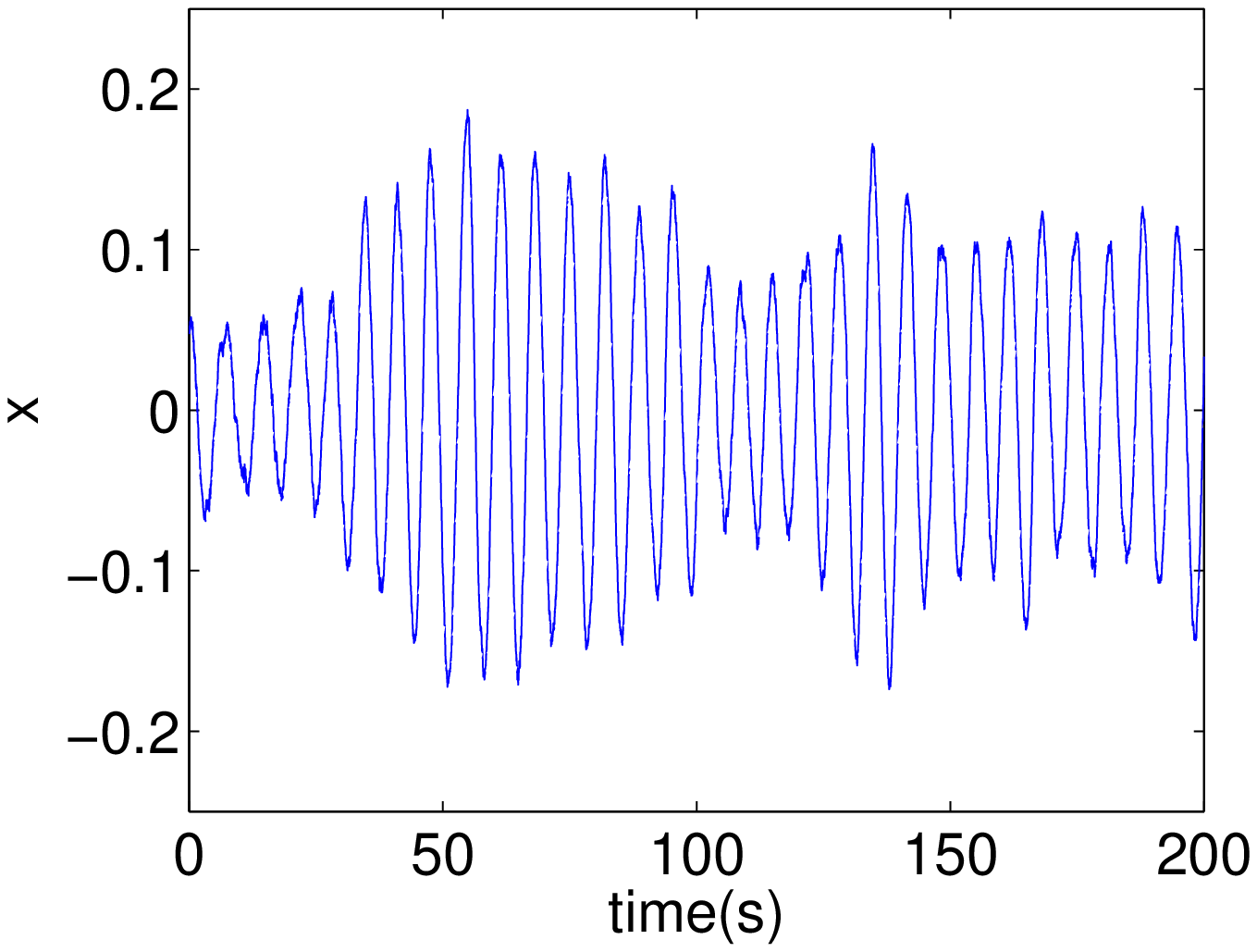}
\caption{}\label{timeseriesLC}
\end{subfigure}
\begin{subfigure}[t]{0.5\linewidth}
\includegraphics[width=1.8in ,keepaspectratio=true,angle=0]{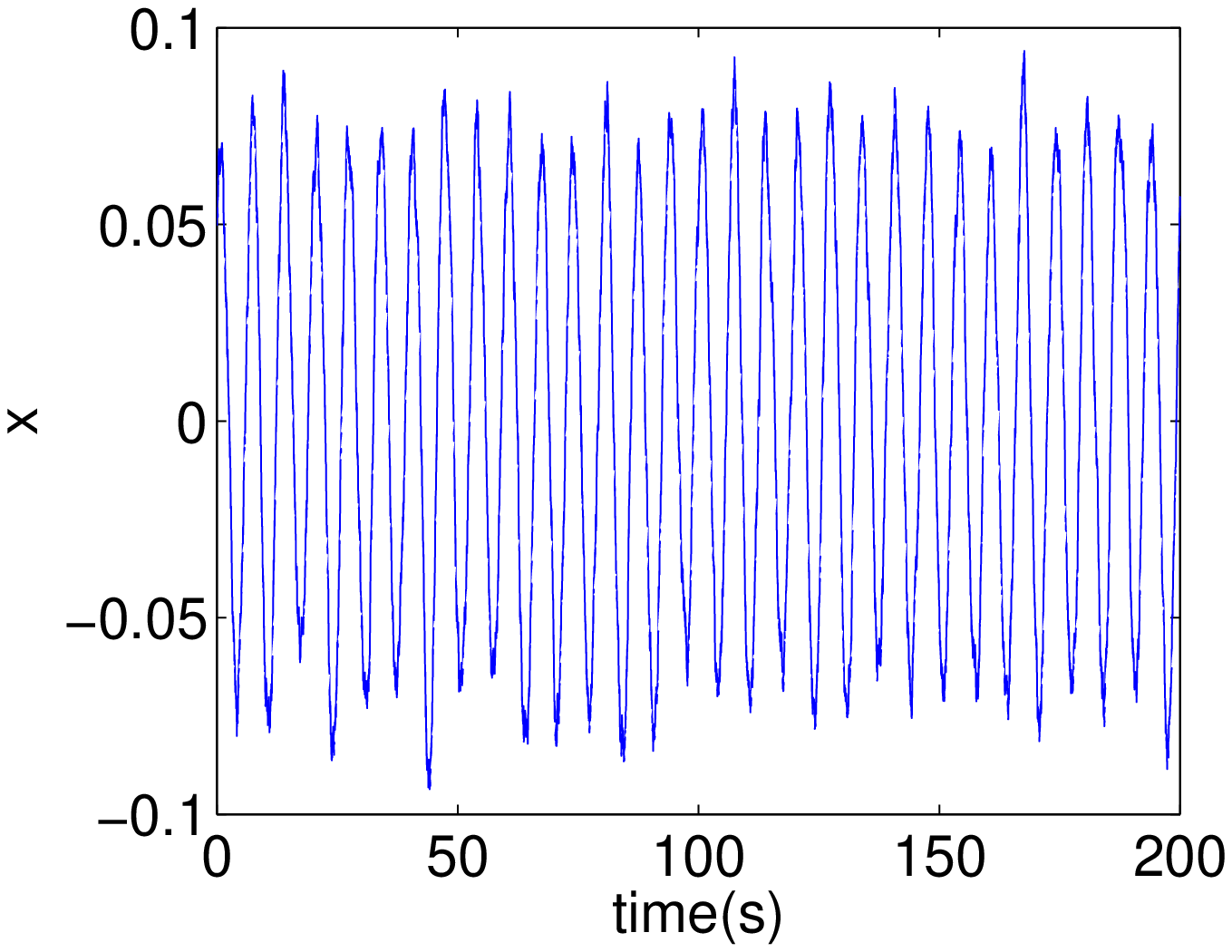}
\caption{}\label{timeseriesForced}
\end{subfigure}
\caption{(a). A sample path of $x$ in a weakly damped system (\ref{WDxy}) with given parameters; (b). A sample path of $x$ in system (\ref{LCxy}) with given parameters, which has a stable limit cycle; (c). A sample path of $x$ in system (\ref{Forcedxy}) with given parameters, which has a forced oscillation.}
\end{figure}

Let $z=x+iy$, $\xi=\xi_x+i\xi_y$, and represent (\ref{forced}) in Cartesian coordinate, we have:
\begin{equation}\label{Forcedxy}
\left(\begin{array}{c}\dot{x}\\\dot{y}\end{array}\right)=\left(\begin{array}{cc}-\gamma&-\omega_0\\ \omega_0&-\gamma\end{array}\right)\left(\begin{array}{c}{x}\\{y}\end{array}\right)+\left(\begin{array}{c}F\cos(\Omega t) + \sigma \xi_x\\ F\sin(\Omega t) + \sigma \xi_y\end{array}\right)
\end{equation}
Fig. \ref{timeseriesForced} shows a typical sample path of $x$ in system (\ref{forced}) with $\gamma=1$, $\omega_0=0.2\pi$, $F=0.1$, $\Omega=0.3\pi$, and $\sigma=0.01$.

From Fig. \ref{timeseriesWD}-\ref{timeseriesForced}, it is clearly seen that power system oscillations with different mechanisms all look similar to each other from time series data, which makes it hard to diagnose the exact cause and adopt the right control actions. In the next section, a novel method will be presented to identify the mechanisms for sustained oscillations by exploring the statistical signatures of time series data.

\section{statistical signatures of different oscillation mechanisms}\label{sectionstatistical signatures}
In this section, several statistical characteristics will be briefly introduced, and then elaborated analytically for different oscillation mechanisms. It will be shown that time series data of oscillation in different mechanisms exhibit different statistical signatures, based on which a diagnosis method is developed.

Kurtosis is a descriptor of the shape of a probability distribution, and can be regarded as a measure of deviation from Gaussian distribution.
Mathematically, kurtosis is defined as the standardized fourth  moment around the mean of a distribution \cite{Glasserman:book}\cite{Decarlo:article}:
\begin{equation}
K=\frac{\mathrm{E}(x-\mu)^4}{(\mathrm{E}(x-\mu)^2)^2}=\frac{\mu_4}{\Var^2[x]}
\end{equation}
where $\mathrm{E}$ is the expectation operator, $\mu$ is the mean, $\mu_4$ is the fourth moment.
The Gaussian distribution has a kurtosis of 3, so $\bar{K}=K-3$, termed as excess kurtosis, is often used so that the reference Gaussian distribution has a kurtosis of zero. In this paper, we use $\bar{K}$ in the subsequent analysis. For symmetric unimodal distributions, positive excess kurtosis indicates heavy tails and peakedness relative to the normal distribution, while negative excess kurtosis indicates light tails and flatness. In addition, the kurtosis is not affected by the variance since it is scaled with respect to the variance.

The power spectral density (PSD) describes how the strength of a signal is distributed in the frequency domain. By Wiener-Khinchin theorem:
\begin{equation}
S_x(\omega)=\int_{-\infty}^{+\infty}R_{x}(\tau)e^{-j\omega\tau}d\tau
\end{equation}
where $R_{x}(\tau)=\mathrm{E}(x(t)x(t+\tau))$ is the autocorrelation function. For example, periodic signals give peaks at a fundamental frequency and its harmonics, while white noise has a flat PSD.

It will be shown that the kurtosis together with the PSD function can be used to distinguish the different mechanisms of oscillations. Firstly, the kurtosis can be used to distinguish the weakly damped oscillation from the limit cycle and the forced oscillation.
Then, the PSD can be used to differentiate the limit cycle and the forced oscillation.
Detailed analysis is to be presented for each scenario.
\subsection{Weakly Damped Oscillation}
System (\ref{WDxy}) can be represented as:
\begin{equation}
\dot{\bm{q}}=M\bm{q}+\Sigma\bm{\xi_q}
\end{equation}
where $\bm{q}=[x,y]^T$ is an vector Ornstein-Uhlenbeck process, $\bm{\xi_q}=[\xi_x,\xi_y]^T$, $M=\left(\begin{array}{cc}-\gamma&-\omega_0\\ \omega_0&-\gamma\end{array}\right)$, $\Sigma=\left(\begin{array}{cc}\sigma&0\\0&\sigma\end{array}\right)$. Then the PSD function can be calculated \cite{Gardiner:2009}:
\begin{equation}
S_{\bm{q}}(\omega)=\frac{1}{2\pi}(M+i\omega)^{-1}NN^T(M^T-i\omega)^{-1}
\end{equation}
Specifically, the PSD of $x$ is:
\begin{equation}
S_x(\omega)=\frac{\sigma^2(\gamma^2+\omega_0^2+\omega^2)}{2\pi[(\gamma^2+\omega_0^2-\omega^2)^2+4\gamma^2\omega^2]}
\end{equation}
\begin{figure}[!ht]
\centering
\includegraphics[width=2in ,keepaspectratio=true,angle=0]{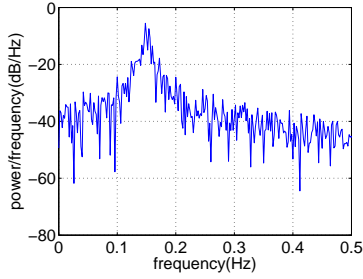}
\caption{Power spectral density of the sample path in Fig. \ref{timeseriesWD}.}\label{WDACPSD}
\end{figure}

Fig. \ref{WDACPSD} shows the PSD of the sample path presented in Fig. \ref{timeseriesWD}.
The power spectrum has a peak around $f_0=0.15\mbox{Hz}$ with a wide bandwidth under the condition that $\gamma\approx0$. It indicates that if the system is weakly damped, the random excitation by the noise may lead to an oscillation near the natural frequency of the system.

Regarding the kurtosis, the process $x(t)$ is Gaussian since all linear functionals of a Gaussian process are also Gaussian. Hence, the kurtosis of $x(t)$ should be approximately zero. Indeed, for the example shown in Fig. \ref{timeseriesWD}, the kurtosis of the sample path is 0.24. 

\subsection{Limit Cycle}
\begin{figure}[!ht]
\centering
\includegraphics[width=2in ,keepaspectratio=true,angle=0]{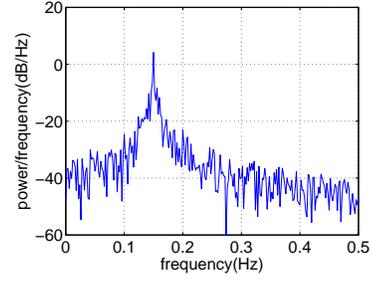}
\caption{Power spectral density of the sample path in Fig. \ref{timeseriesLC}.}\label{LC-ACPSD}
\end{figure}

Consider system (\ref{LCxy}), near the Hopf bifurcating point, the amplitude of the limit cycle grows with $\sqrt{\gamma}$, and the angular frequency is approximately $\omega_h$. The solution can be approximated as:
\begin{equation}
\left(\begin{array}{c}x(t)\\y(t)\end{array}\right)\approx\left(\begin{array}{c}(\sqrt{\gamma}+p(t))\cos(\phi(t))\\(\sqrt{\gamma}+p(t))\sin(\phi(t))\end{array}\right)
\end{equation}
where $\dot{p}=-2\gamma p+\sigma_{p}\xi_{p}$ is an Ornstein-Uhlenbeck process independent of $\phi$, $\phi(t)=\omega_h t+\sigma_{\phi}\xi_{\phi}$ is a Brownian motion with deterministic drift \cite{Louca:2015}. The PSD of $x(t)$ can be represented as\cite{d'Onofrio:2013}:
\begin{equation}
S_x(\omega)=\gamma F(\sigma_\phi^2,\omega_h,\omega)+\frac{\sigma_{p}^2}{4\gamma}F(\sigma_\phi^2+4\gamma,\omega_h,\omega)
\end{equation}
where $F$ is the PSD of $\cos(\phi(t))$:
\begin{equation}
F(\sigma_\phi^2,\omega_h,\omega)=\frac{\sigma^2_\phi(\omega^2+\omega_h^2+\sigma_{\phi}^2/4)}{4\pi(\omega^2-\omega_h^2-\sigma_\phi^2/4)^2+\sigma_\phi^4\omega^2}
\end{equation}
The phase noise leads to a shift of the spectral peak, and even for very stable limit cycles ($\gamma\gg\sigma_{\phi}$), the power spectrum has a non-vanishing bandwidth, leading to the temporal decoherence which is known as ``jitter" in electronic signal theory \cite{Louca:2014}\cite{Demir:2000}. Fig. \ref{LC-ACPSD} shows the PSD of the sample path in Fig. \ref{timeseriesLC}, which has a peak around 0.15Hz with a wide bandwidth.

Even though the weakly damped oscillation and the limit cycle have similar power spectrums, they have different properties with respect to the kurtosis. The kurtosis of $x(t)=(\sqrt{\gamma}+p(t))\cos(\phi(t))$ is given below, refer to Appendix \ref{kurtosisderivation} for detailed derivation.
\begin{equation}
\mathrm{Kurt}[x(t)]=-\frac{3[(\gamma-\Var[p(t)])^2-2(\Var^2[p(t)])]}{2(\gamma+\Var[p(t)])^2}\label{LCkurtosis}
\end{equation}
From (\ref{LCkurtosis}), we have that if $\frac{\gamma}{\Var[p(t)])}=k>(1+\sqrt{2})$, the kurtosis is negative; as $k$ grows, i.e. the amplitude of the limit cycle is more and more larger than the noise intensity $\Var[p(t)]$, the kurtosis gets more and more closer to $-\frac{3}{2}$, which makes $x(t)$ more distinguishable from Gaussian distribution, and thus from the weakly damped oscillation. For the example shown in Fig. \ref{timeseriesLC}, the kurtosis is -1.05, which is much farther away from zero compared with the weakly damped oscillation.
\subsection{Forced Oscillation}
We next consider system (\ref{Forcedxy}) whose solution can be represented as:
\begin{equation}
\dot{\bm{q}}=\bm{q_0}+\bm{q_1}
\end{equation}
where $\bm{q_0}=[x_0,y_0]^T$ satisfies
\begin{equation}
\dot{\bm{q_0}}=M\bm{q_0}+\left(\begin{array}{c}F\cos(\Omega t)\\F\sin(\Omega t)\end{array}\right)
\end{equation}
and $\bm{q_1}=[x_1,y_1]^T$ is a vector Ornstein-Uhlenbeck process:
\begin{equation}
\dot{\bm{q_1}}=M\bm{q_1}+\Sigma\bm{\xi_q}
\end{equation}

We consider the real part $x(t)$, the stationary solution (i.e., $t\to +\infty$) of $x$ can be represented as:
$x(t)=\rho\cos(\Omega t)+x_1(t)$, with the PSD function:
\begin{eqnarray}\label{forcedPSD}
S_x(\omega)&=&\frac{1}{2}\rho^2[\delta(\omega_0-\Omega)+\delta(\omega_0+\Omega)]\\
&&+\frac{\sigma^2(\gamma^2+\omega_0^2+\omega^2)}{2\pi(\gamma^2+\omega_0^2-\omega^2)^2+4\gamma^2\omega^2}\nonumber
\end{eqnarray}

\begin{figure}[!ht]
\centering
\includegraphics[width=2in ,keepaspectratio=true,angle=0]{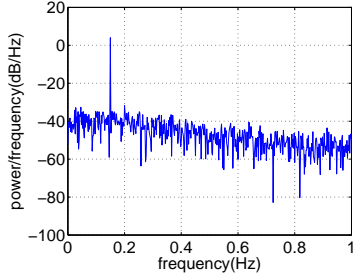}
\caption{Power spectral density of the sample path in Fig. \ref{timeseriesForced}.}\label{forcedACPSD}
\end{figure}

Fig. \ref{forcedACPSD} presents the PSD of the sample path shown in Fig. \ref{timeseriesForced}. It demonstrates that in a well-damped system, the PSD function has a thin spike at the frequency $\frac{\Omega}{2\pi}$ corresponding to the $\delta$ function in (\ref{forcedPSD}).

Compare Fig. \ref{forcedACPSD} with Fig. \ref{LC-ACPSD} and Fig. \ref{WDACPSD}, the forced oscillation is characterized by a thin spike in the power spectrum, which is distinct from the other oscillation mechanisms. In practice, the width of the spike will be determined by the decoherence time of the external forcing, so it will be narrow for truly periodic load variations.

Furthermore, the kurtosis of $x(t)=\rho\cos(\Omega t)+x_1(t)$ can be represented as below, whose detailed derivation is given in Appendix \ref{kurtosisderivation}.
\begin{equation}
\mathrm{Kurt}[x(t)]=\frac{\E[(x(t)-\rho)^4]}{\Var^2[x(t)]}-3=-\frac{3}{2}\frac{1}{(1+2\frac{\Var[x_1(t)]}{\rho^2})^2}\label{forcedkurtosis}
\end{equation}
(\ref{forcedkurtosis}) implies that the kurtosis of $x(t)$ is negative, which means that the distribution of $x(t)$ has light tails and flatness compared with Gaussian distribution. In addition, $\mathrm{Kurt}[x(t)]\geq-\frac{3}{2}$, and the absolute value of $\mathrm{Kurt}[x(t)]$ depends on $\frac{\Var[x_1(t)]}{\rho^2}$.
If the noise intensity of $x_1(t)$ is much less than the amplitude of the oscillation $\rho$, the kurtosis will be far away from zero and close to $-\frac{3}{2}$, which makes the distribution of $x(t)$ distinguishable from Gaussian distribution. Indeed, for the example shown in Fig. \ref{timeseriesForced}, the kurtosis is -1.44.

\begin{figure}[!ht]
\centering
\includegraphics[width=3.7in ,keepaspectratio=true,angle=0]{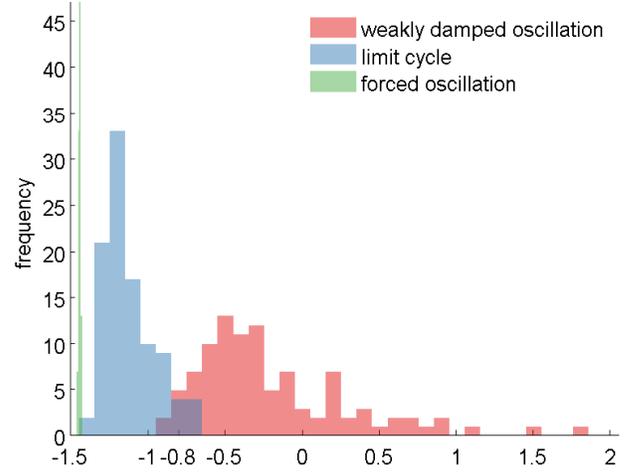}
\caption{Histograms of kurtosis value for all mechanisms.}\label{montecarlo}
\end{figure}

Monte Carlo simulations are also done to demonstrate the characteristic of kurtosis for each mechanism. 500s simulation has been run for 100 times for each mechanism, and histograms of kurtosis value are shown in Fig. \ref{montecarlo}. The widely spread of kurtosis value makes different oscillation mechanisms readily distinguishable. Additionally, 90\% confidence levels have been calculated. 90\% confidence interval of the kurtosis for weakly damped oscillation is [-0.68, 0.56], 90\% confidence interval of the kurtosis for limit cycle is [-1.29, -0.88], and that of the kurtosis for forced oscillation is [-1.45 -1.43]. The histograms and confidence intervals have demonstrated the pronounced distinguishability of kurtosis.

From the above analysis, we see that different oscillation mechanisms exhibit different properties regarding the kurtosis and the PSD. Specifically, the kurtosis can be used to distinguish the weakly damped oscillation, which has approximately zero kurtosis, from the forced oscillation and the limit cycle, which have negative kurtosis. Furthermore, the PSD can be used to distinguish the limit cycle which has a wide bandwidth, and the forced oscillation which has a thin spike.
\begin{figure}[!ht]
\centering
\includegraphics[width=3in ,keepaspectratio=true,angle=0]{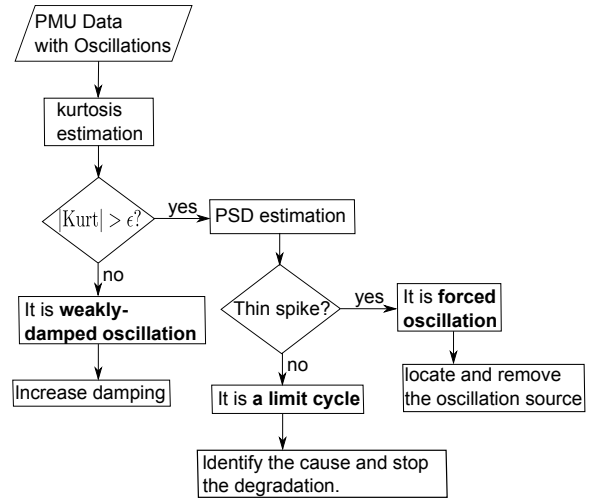}
\caption{Flowchart of the algorithm to diagnose sustained oscillations.}\label{flowchart}
\end{figure}
We hence propose an algorithm to diagnose the sustained oscillation causes as shown in Fig. \ref{flowchart}. The threshold $\ee$ has to be decided by considering noise intensity, confidence interval, system features and so forth in practical implementation; the detection of thin spike also depends on the tradeoff between noise intensity and forcing strength. Further studies are needed to improve the decision procedures in this algorithm.
Besides, further investigations about control design are needed, while some preliminary results are to be presented in the following case studies, which is to locate the oscillation source or problematic components via the kurtosis.

\section{Case Studies}\label{sectioncasestudies}
\begin{figure}[!ht]
\centering
\begin{subfigure}[t]{0.5\linewidth}
\includegraphics[width=1.8in ,keepaspectratio=true,angle=0]{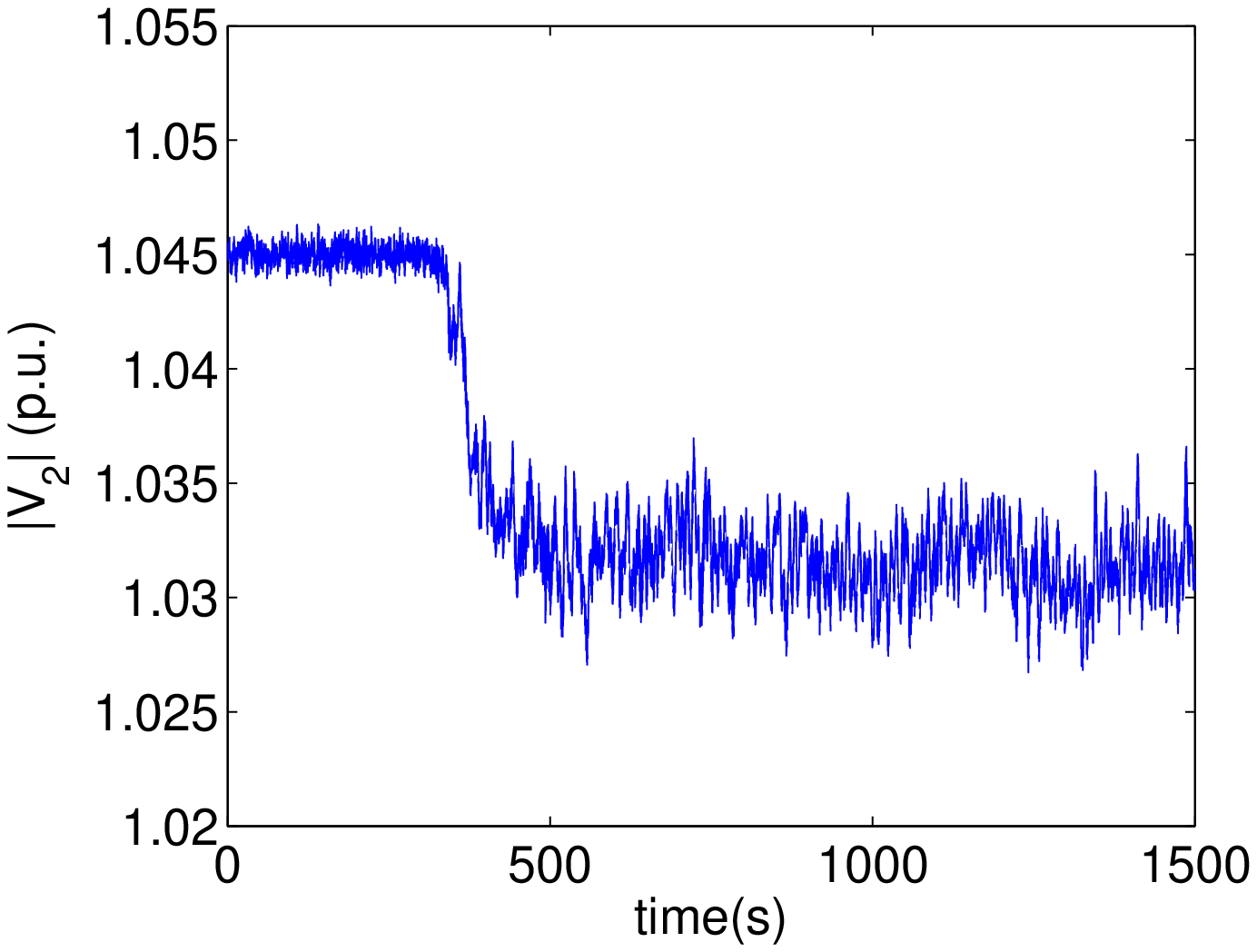}
\caption{Voltage magnitude at Bus 2}\label{voltage-a}
\end{subfigure}%
\begin{subfigure}[t]{0.5\linewidth}
\includegraphics[width=1.8in ,keepaspectratio=true,angle=0]{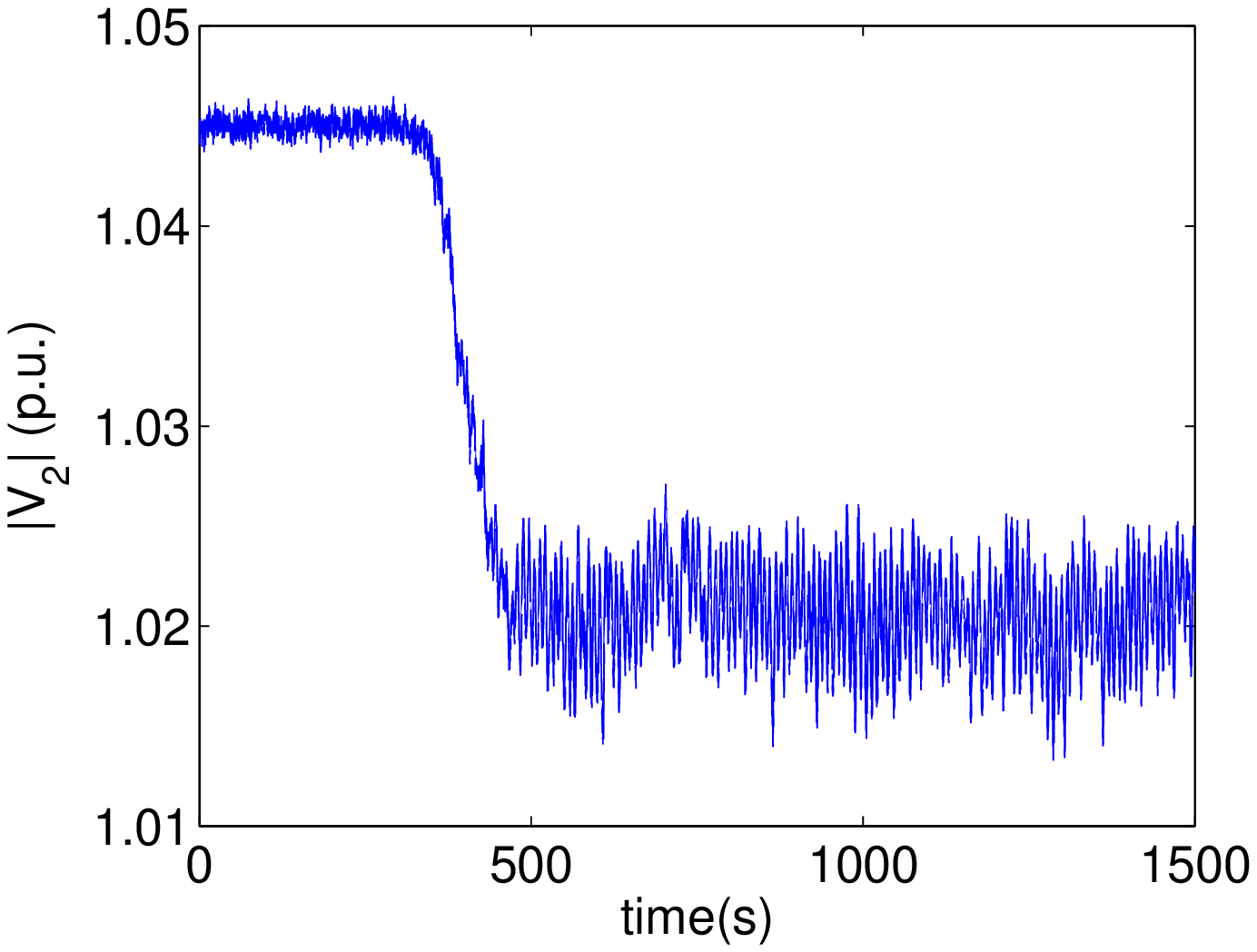}
\caption{Voltage magnitude at Bus 2}\label{voltage-b}
\end{subfigure}
\begin{subfigure}[t]{0.5\linewidth}
\includegraphics[width=1.8in ,keepaspectratio=true,angle=0]{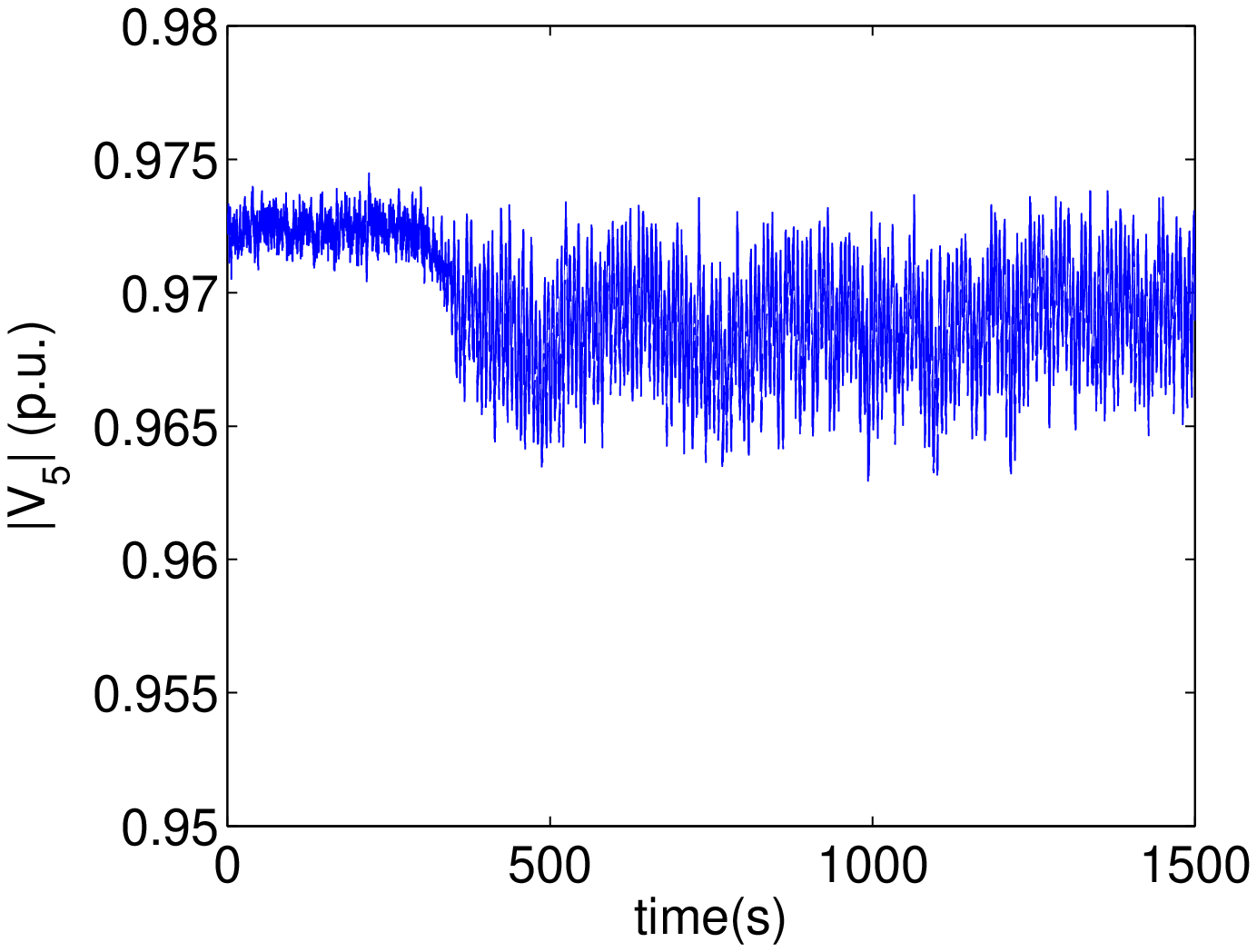}
\caption{Voltage magnitude at Bus 5}\label{voltage-c}
\end{subfigure}
\caption{One specific voltage magnitude in three cases.}\label{voltage}
\end{figure}

Fig. \ref{voltage} shows three different oscillation scenarios in 14-bus systems, from which the exact mechanism of each scenario can be hardly identified. In this section, we apply the proposed diagnostic method to identify the oscillation mechanism for each scenario using the simulated PMU data. Note that all the time series data is obtained from time-domain simulation of three different 14-bus systems which are modified from the IEEE 14-bus benchmark system. Stochastic loads are added to the systems. Particularly, all load fluctuations follow the Ornstein-Uhlenbeck process, with $\sigma=0.01$ in (\ref{sto eqn}). Besides, we use the exponential recovery load which has been widely used in studying voltage stability \cite{Kundur:book}\cite{Cutsem:book} and can be modelled as follows:
\begin{eqnarray}
P&=&kP_0(\frac{V}{V_0})^\alpha\\
Q&=&kQ_0(\frac{V}{V_0})^\beta
\end{eqnarray}
where $k$ is a dimensionless demand variable, $V_0$ is the reference voltage, and $\alpha$ and $\beta$ are active and reactive power exponents depending on the type of load \cite{Cutsem:book}\cite{Milano:article}.  All simulations were performed on PSAT-2.1.8 \cite{Milano:article}. Parameter values of the base case are given in Appendix \ref{appendixbasecase}, and all test systems are posted online: https://github.com/xiaozhew/Test-Systems-.
\subsection{Case I}
From the time series data shown in Fig. \ref{voltage-a}, we see that the system is slowly evolving between 300s and 700s probably due to changing parameters. Given that the new steady state (between 700s to 1500s) suffers from oscillation, we intend to figure out the exact mechanism of the oscillation.

\begin{figure}[!ht]
\centering
\begin{subfigure}[t]{0.5\linewidth}
\includegraphics[width=1.8in ,keepaspectratio=true,angle=0]{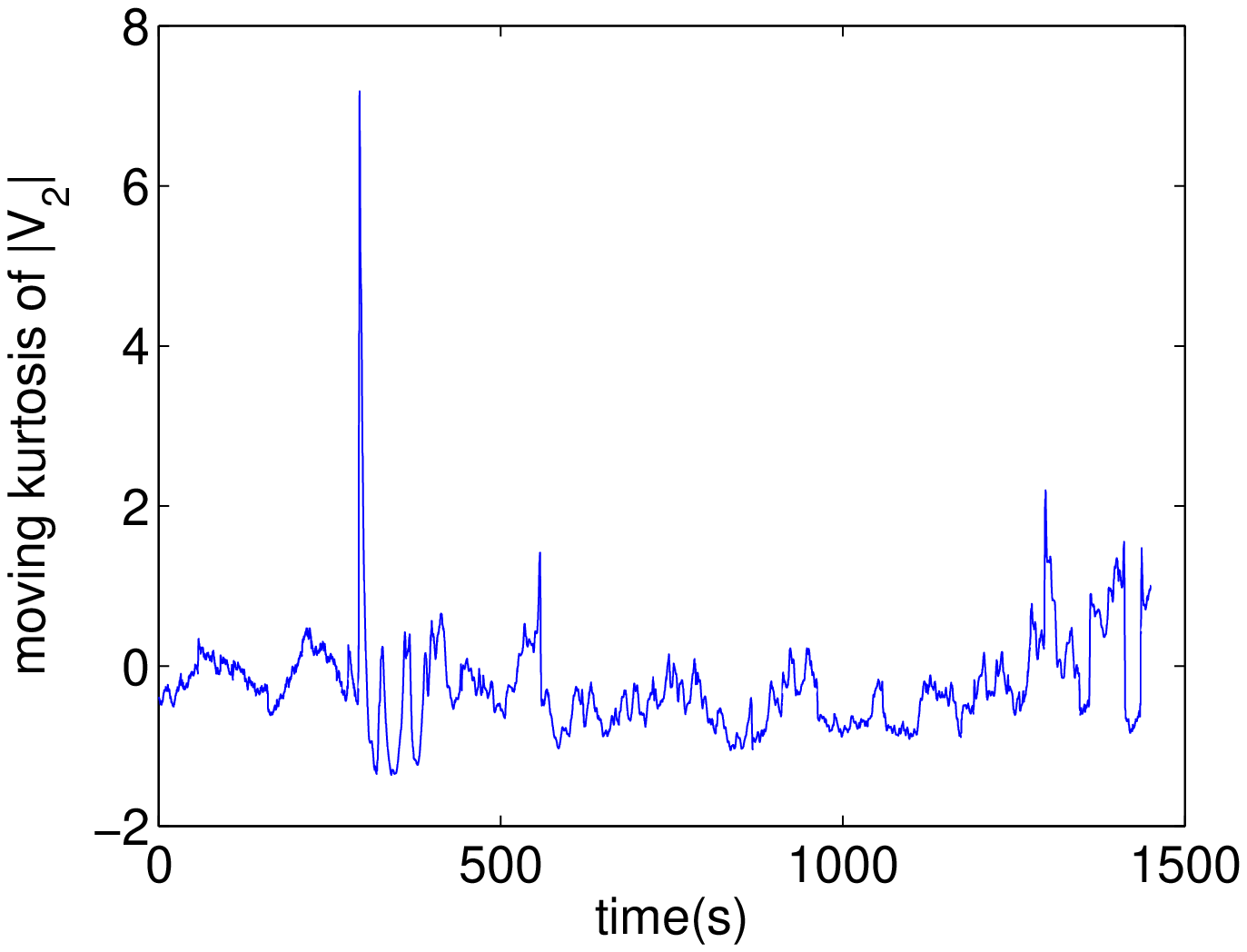}
\caption{}\label{kurtosisV2WD}
\end{subfigure}%
\begin{subfigure}[t]{0.5\linewidth}
\includegraphics[width=1.8in ,keepaspectratio=true,angle=0]{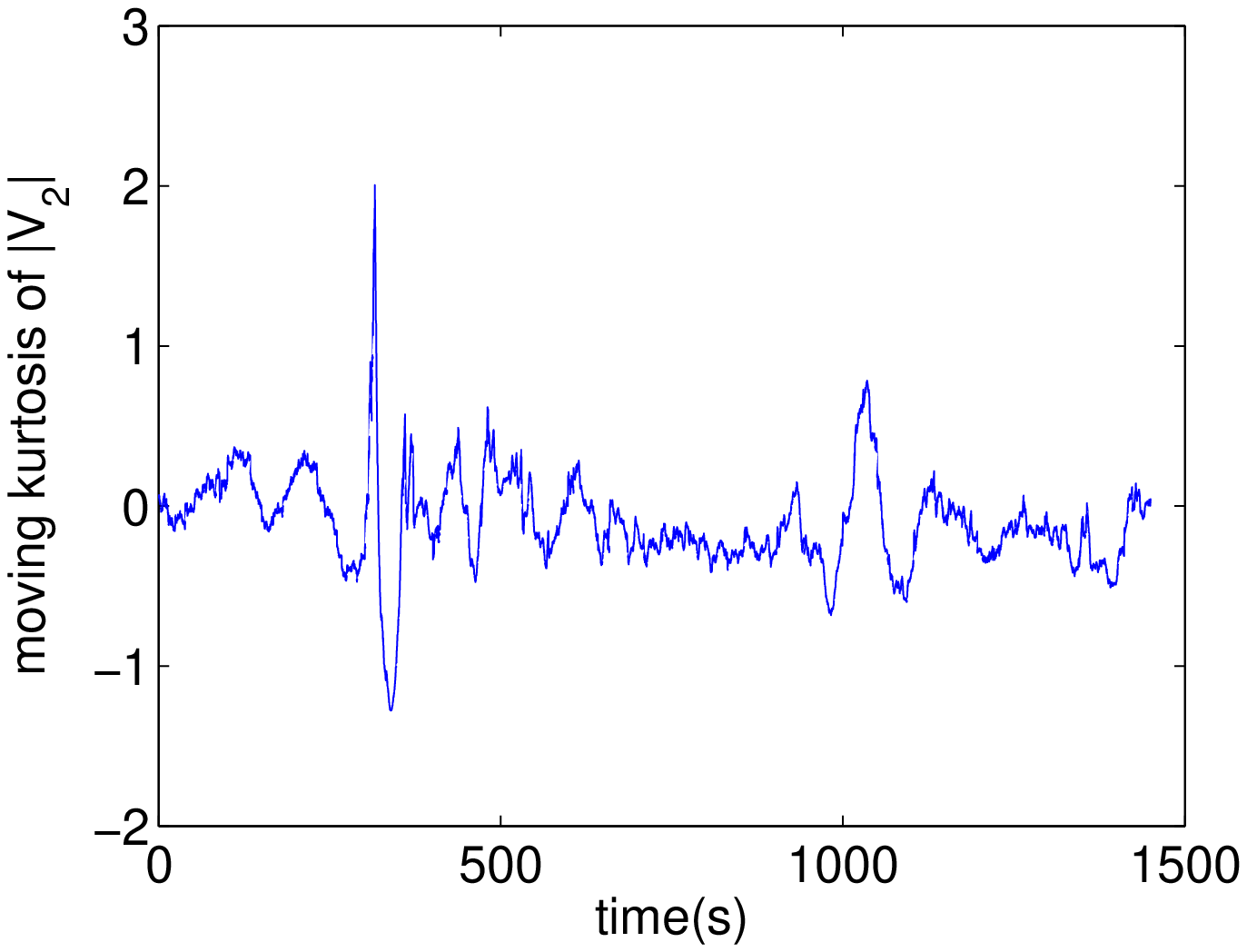}
\caption{}\label{kurtosisV2WD-withnoise}
\end{subfigure}
\caption{(a). The moving kurtosis of the voltage magnitude shown in Fig. \ref{voltage-a} without measurement noise.; (b). The moving kurtosis of the voltage magnitude under the influence of measurement noise.}
\end{figure}

Following the algorithm shown in Fig. \ref{flowchart}, we firstly estimate the kurtosis of the time series data in Fig. \ref{voltage-a}. The moving kurtosis of the time series data with a window size of 50s is presented in Fig. \ref{kurtosisV2WD}. The peak in the figure corresponds to the transient dynamics due to changing parameters, while the kurtosis before and after the variation doesn't change much. In fact, the kurtosis of the new steady state (between 700s to 1500s) is 0.006, which indicates that the random process is still approximately Gaussian and the nonlinearity doesn't outstand. Therefore, we conclude that the sustained oscillation is due to weak damping. It implies that as the system evolves, the principal eigenvalue is getting closer to the imaginary axis, or in other words, the systems is getting closer to its stability boundary.

We also examine the impact of measurement noise on the proposed oscillation diagnostic method. Measurement noise may wash out useful information and affect the observed statistics, and thus affect the detection and decision. To address this issue, the band-pass filter was applied to filter out measurement noise in \cite{Hines:2015}. It has been shown that the standard deviation (STD) of the measurement noise after filtering can be reduced to the order of $10^{-3}$ if the STD of the original measurement noise is $10^{-2}$. Following \cite{Hines:2015}, we apply a white Gaussian noise with $\mbox{STD}=10^{-3}$ as the measurement noise to the simulated PMU data, and obtain the corresponding kurtosis shown in Fig. \ref{kurtosisV2WD-withnoise}. The estimated kurtosis of the new steady state (between 700s to 1500s) is -0.12. Even though the measurement noise deteriorates the statistic, it is still obvious from Fig. \ref{kurtosisV2WD-withnoise} that the kurtosis is close to zero, and thus system has weakly damped oscillation.

The actual situation is that the variation of system voltages is a result of increasing loads. Starting from 300s, the exponential recovery loads at Bus 5 and Bus 12 increase gradually, and by 380s, both loads grow by 8\% and stop increasing afterwards. The increasing loads require more power support from the generators, making the field currents of generators closer to their limits. As the power required by the loads gets closer to the limit of the generators and the transmission network, the system is pushed closer to its voltage stability boundary, which leads to the weak damping of the complete power system. In order to maintain the stable operation of the system, damping of the system needs to be increased possibly by tuning the parameters of the excitation system.

From this example, we see that the proposed oscillation diagnostic method is able to accurately distinguish the weakly damped oscillation from the other mechanisms, and provide important guidance to further design the right control actions.

\subsection{Case II}
In the second case, the time series data in Fig. \ref{voltage-b} performs similarly as the previous case. Given that the new steady state exhibits oscillations after a transient period between 300s to 700s, we want to diagnose the exact oscillation mechanism.

\begin{figure}[!ht]
\centering
\begin{subfigure}[t]{0.5\linewidth}
\includegraphics[width=1.8in ,keepaspectratio=true,angle=0]{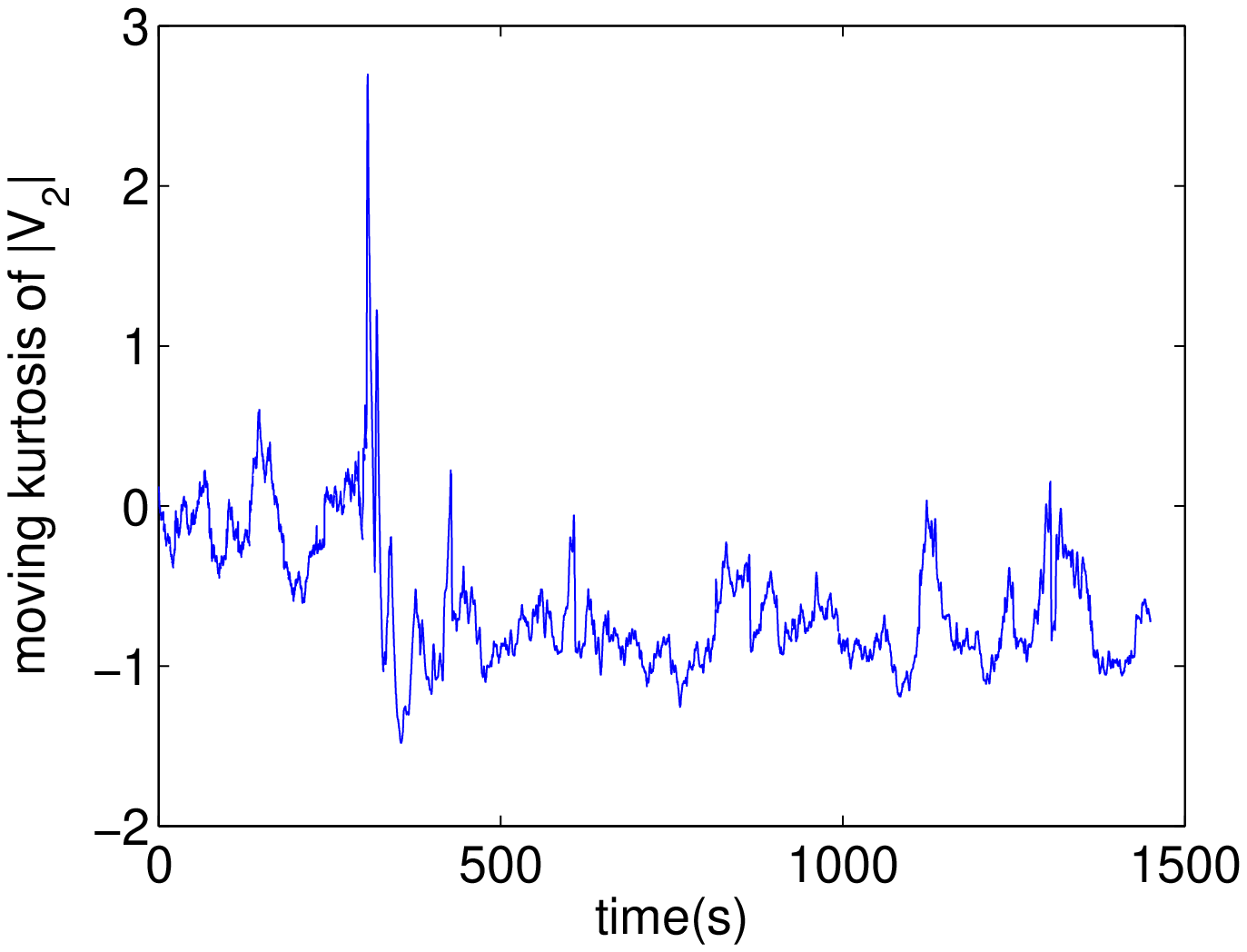}
\caption{}\label{kurtosisV2LC}
\end{subfigure}%
\begin{subfigure}[t]{0.5\linewidth}
\includegraphics[width=1.8in ,keepaspectratio=true,angle=0]{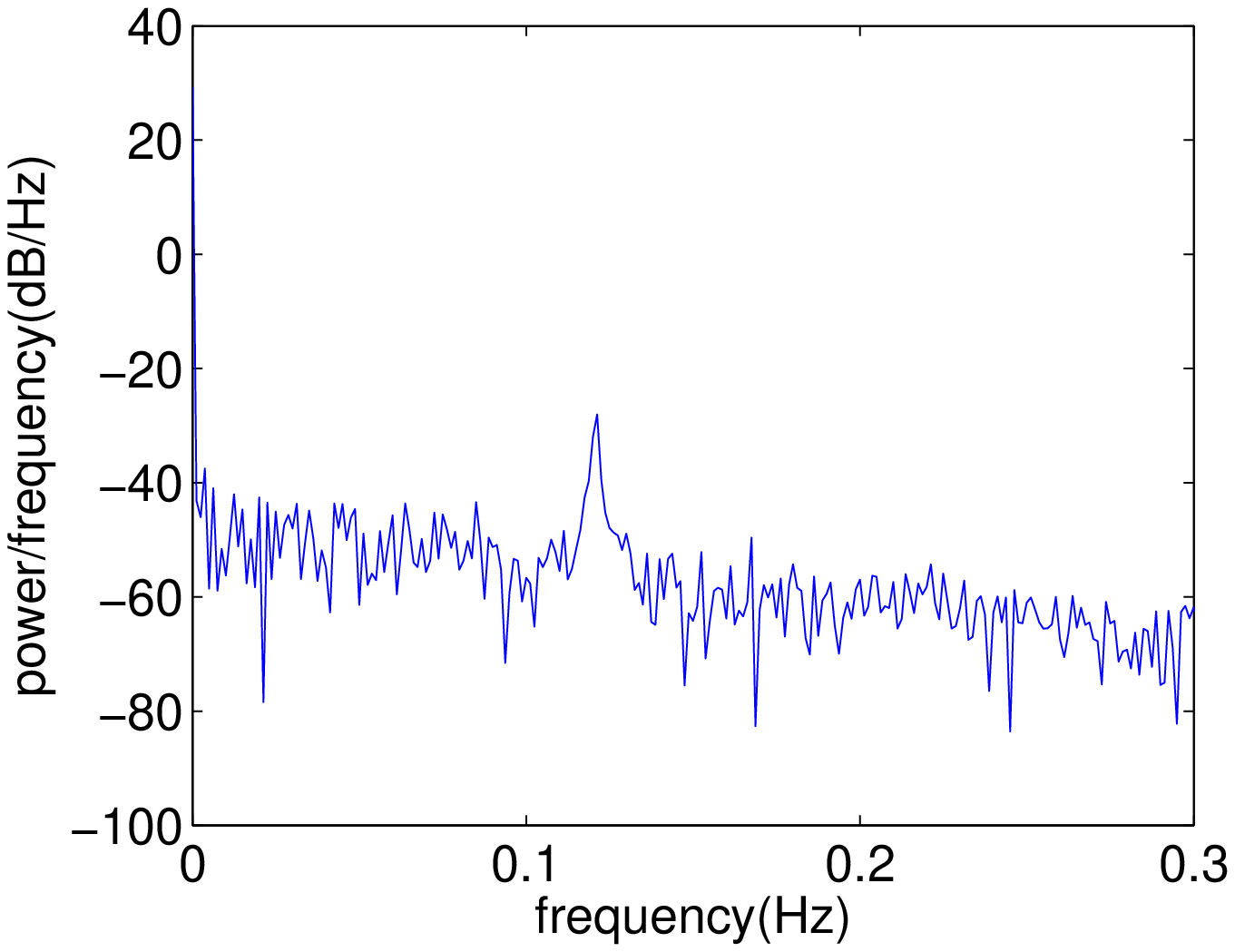}
\caption{}\label{ACPSDV2LC}
\end{subfigure}
\caption{(a). The moving kurtosis of the voltage magnitude shown in Fig. \ref{voltage-b}; (b). The power spectral density of the steady-state time series data between 700s and 1500s in Fig. \ref{voltage-b}.}
\end{figure}

By the proposed method, we firstly estimate the kurtosis to discern the weakly damped oscillation from the other mechanisms. Fig. \ref{kurtosisV2LC} shows the moving kurtosis of the data in Fig. \ref{voltage-b}, from which we can observe a significant change of the kurtosis before and after the variation. In fact, the kurtosis of the new steady-state time series data between 700s and 1500s is -0.58, which implies that the stochastic process has already deviated from Gaussian process. Hence, the weakly damped oscillation has been eliminated from all the mechanisms in this scenario.
Next, we estimate the PSD of the steady-state time series data between 700s and 1500s to differentiate the forced oscillation and the limit cycle. As presented in Fig. \ref{ACPSDV2LC}, there is a clear peak around 0.12Hz in the power spectrum with a wide bandwidth. Hence, we conclude that the system has passed the Hopf Bifurcation point. The original equilibrium point has lost it stability and a stable limit cycle has emerged, which may be an early warning sign of voltage collapse.

\begin{figure}[!ht]
\centering
\begin{subfigure}[t]{0.5\linewidth}
\includegraphics[width=1.8in ,keepaspectratio=true,angle=0]{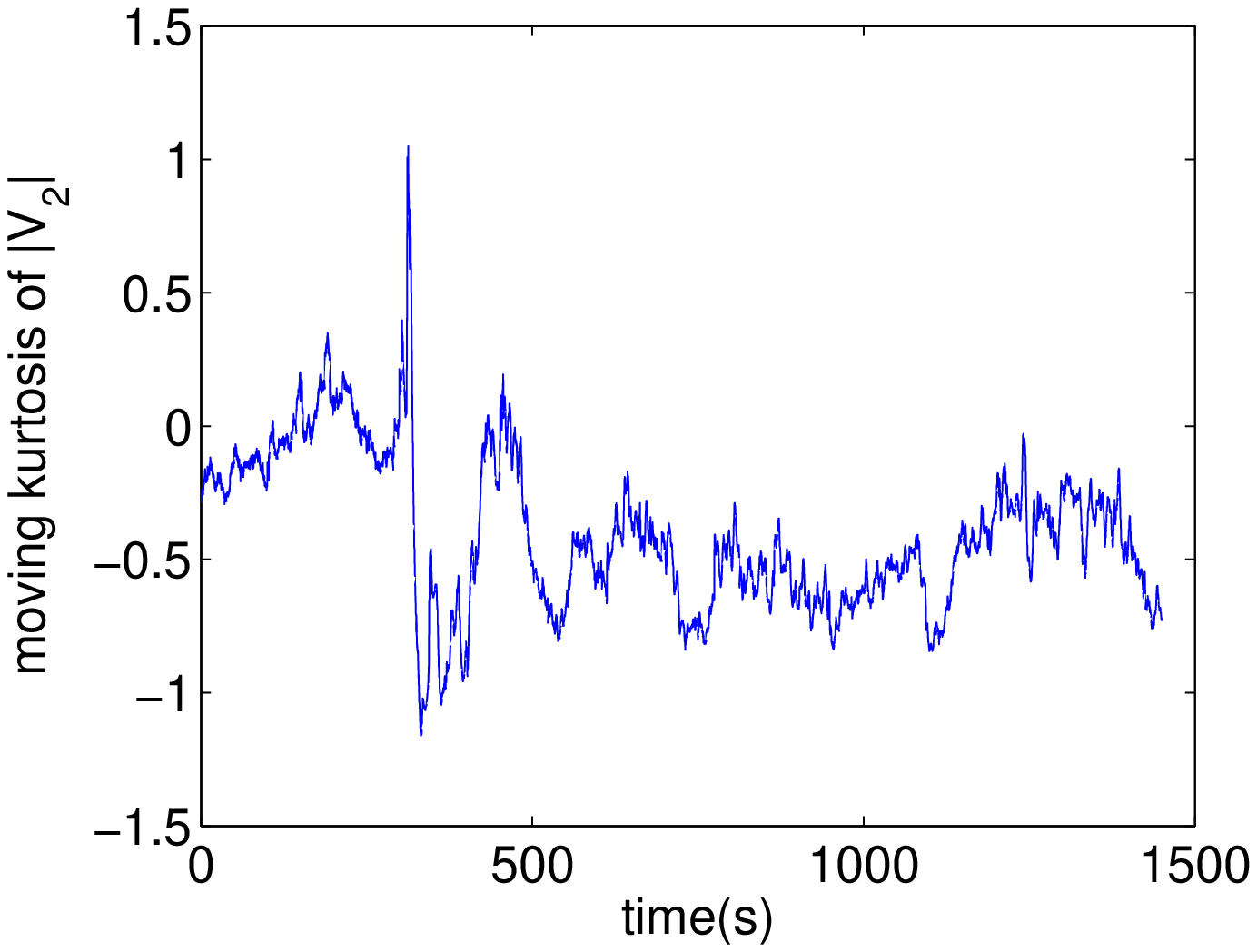}
\caption{}\label{kurtosisV2LC-noise}
\end{subfigure}%
\begin{subfigure}[t]{0.5\linewidth}
\includegraphics[width=1.8in ,keepaspectratio=true,angle=0]{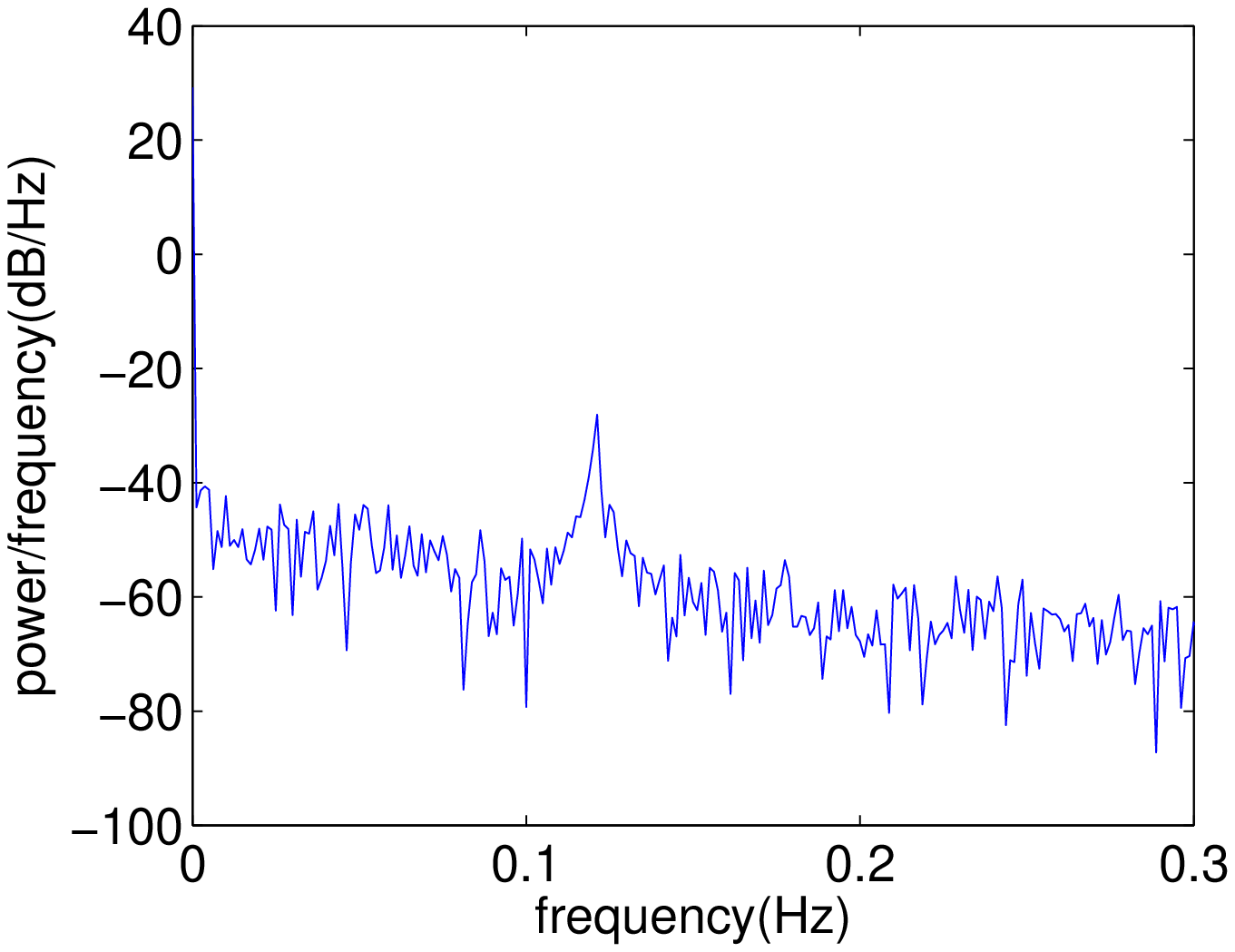}
\caption{}\label{ACPSDV2LC-noise}
\end{subfigure}
\caption{(a). The moving kurtosis of the voltage magnitude with measurement noise; (b). The power spectral density of the steady-state time series data between 700s and 1500s with measurement noise.}
\end{figure}

Similarly, we also incorporates measurement noise in this case. We apply a white Gaussian noise with $\mbox{STD}=10^{-3}$ as the measurement noise to the simulated PMU data, and obtain the corresponding moving kurtosis and PSD as shown in Fig. \ref{kurtosisV2LC-noise}-\ref{ACPSDV2LC-noise}. It is observed that both the kurtosis and the PSD perform similarly as the case without measurement noise. Particularly, the moving kurtosis of the new steady-state time series data between 700s and 1500s is -0.52, and the power spectrum has a clear peak around 0.12Hz with a wide bandwidth. Regarding the SNR, the amplitude of the limit cycle in the subspace of $|V_2|$ is $5\times 10^{-4}$ p.u., and variance of measurement noise is $10^{-6}$, hence the SNR is approximately $-9.03$ dB. In practice, both load fluctuations and electromechanical excitations act as an effective noise, thus making the SNR even lower. These
effects are considered separately from the measurement noise in our study.

In fact, the oscillation in this case is a result of continuously increasing the loads after 380s in \textit{Case I}. The exponential recovery loads at Bus 5 and Bus 12 start to increase at 300s, grow by 12\% at 420s, and stop increasing afterwards. The increasing loads result in the oscillation of over excitation limiter (OXL) for the generator at Bus 2, which further excites the fast variables of AVRs. The effect of OXL to voltage stability has been discussed in extensive literatures \cite{Kundur:book}\cite{Cutsem:book}\cite{Vu:article}-\cite{Wangxz:PSCC}. In this case, the competing effect between load dynamics and OXLs finally leads to the voltage instability. The power system passes the Hopf bifurcation point around 400s when a stable limit cycle is born.

\begin{figure}[!ht]
\centering
\includegraphics[width=3.7in ,keepaspectratio=true,angle=0]{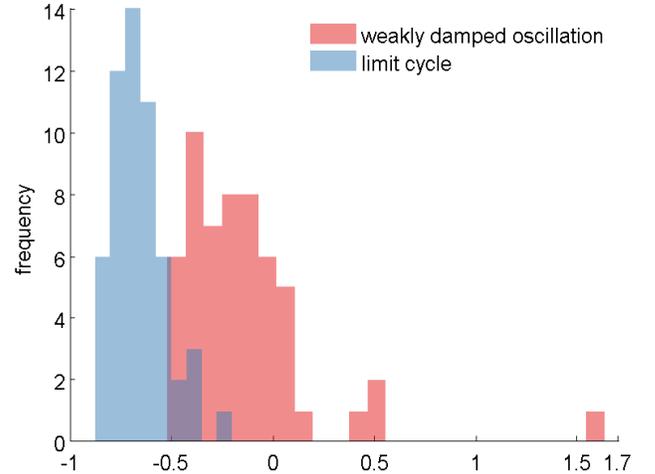}
\caption{\textcolor{blue}{Histograms of kurtosis value for all mechanisms.}}\label{montecarlopowersystem}
\end{figure}

To show the characteristics of kurtosis before and after Hopf bifurcation in this power system, Monte Carlo simulations have been done for both \textit{Case I} and \textit{Case II}.  The histograms in Fig. \ref{montecarlopowersystem} are plotted according to 55 360s samples of $|V_2|$ for both cases. The 90\% confidence interval of kurtosis for \textit{Case I} (i.e., weakly damped oscillation) is [-0.44, 0.10], and the 90\% confidence interval of kurtosis for \textit{Case II} (i.e., limit cycle) is [-0.82, -0.46]. The confidence intervals do not overlap with each other, which makes the decision procedure very easy. These results further validate the ability of kurtosis to distinguish between weakly damped oscillation and limit cycle.

This example shows that the proposed oscillation diagnosis method is able to identify the limit cycle from all the mechanisms even when the SNR is low, and thus provides significant guidance to adopt the right control strategy in time. In this example, load increasing needs to be stopped to avoid further voltage degradation or even voltage collapse.

We further observed that the kurtosis of real-time data may also help locate the oscillation source or the problematic components, because the absolute value of the kurtosis indicates the amplitude of oscillation under fixed noise intensity as shown in (\ref{LCkurtosis}). By comparing the kurtosis of the voltage magnitudes at different buses which can be achieved from PMU data, the problematic components like the most stressed generator may be identified. In this example, the kurtosis of each voltage magnitude is shown in Table \ref{TablekurtosisLC}. For illustration purpose, we do not consider measurement noise here. We see that the kurtosis of the voltage magnitude at Bus 2 has the maximum absolute value which indicates that the amplitude of voltage oscillation is larger at Bus 2 compared with other buses. The oscillation with larger amplitude implies more severe problem and potential oscillation source, which is true in this case. The activation of OXL of the generator at Bus 2 results in the excitation of other fast variables of AVRs, and finally leads to the voltage oscillation issue. Additionally, as the power support for the load at Bus 5 is mainly from the generator at Bus 2, the load at Bus 5 needs to stop increasing in order to avoid further voltage degradation; parameters of the excitation system of the generator at Bus 2 also need to be tuned to suppress the oscillation.

\begin{table}[!ht]
\centering
\normalsize{
\begin{tabular}{|c|c|c|c|}
\hline
Bus 1&-0.34& Bus 8&-0.22\\
\hline
Bus 2&\textbf{-0.54}& Bus 9&-0.17\\
\hline
Bus 3&-0.05& Bus 10&-0.18\\
\hline
Bus 4&-0.23& Bus 11&-0.20\\
\hline
Bus 5&-0.22& Bus 12&-0.23\\
\hline
Bus 6&-0.19& Bus 13&-0.21\\
\hline
Bus 7&-0.23& Bus 14&-0.13\\
\hline
\end{tabular}
}
\caption{The kurtosis of voltage magnitudes at different buses for the new steady state (between 700s and 1500s) in \textit{Case II}.}\label{TablekurtosisLC}
\end{table}

\subsection{Case III}
\begin{figure}[!ht]
\centering
\begin{subfigure}[t]{0.5\linewidth}
\includegraphics[width=1.8in ,keepaspectratio=true,angle=0]{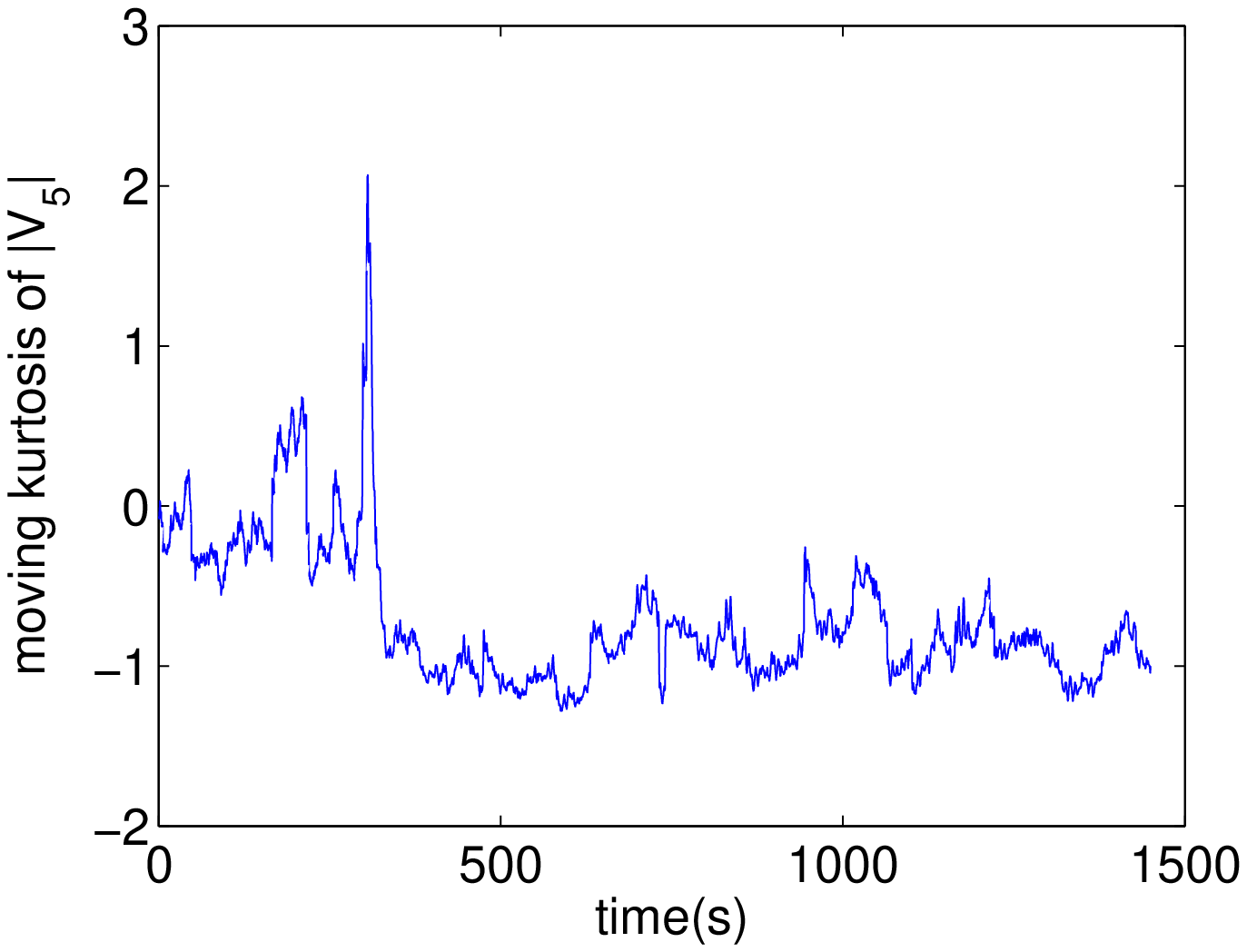}
\caption{}\label{kurtosisV5forced}
\end{subfigure}%
\begin{subfigure}[t]{0.5\linewidth}
\includegraphics[width=1.8in ,keepaspectratio=true,angle=0]{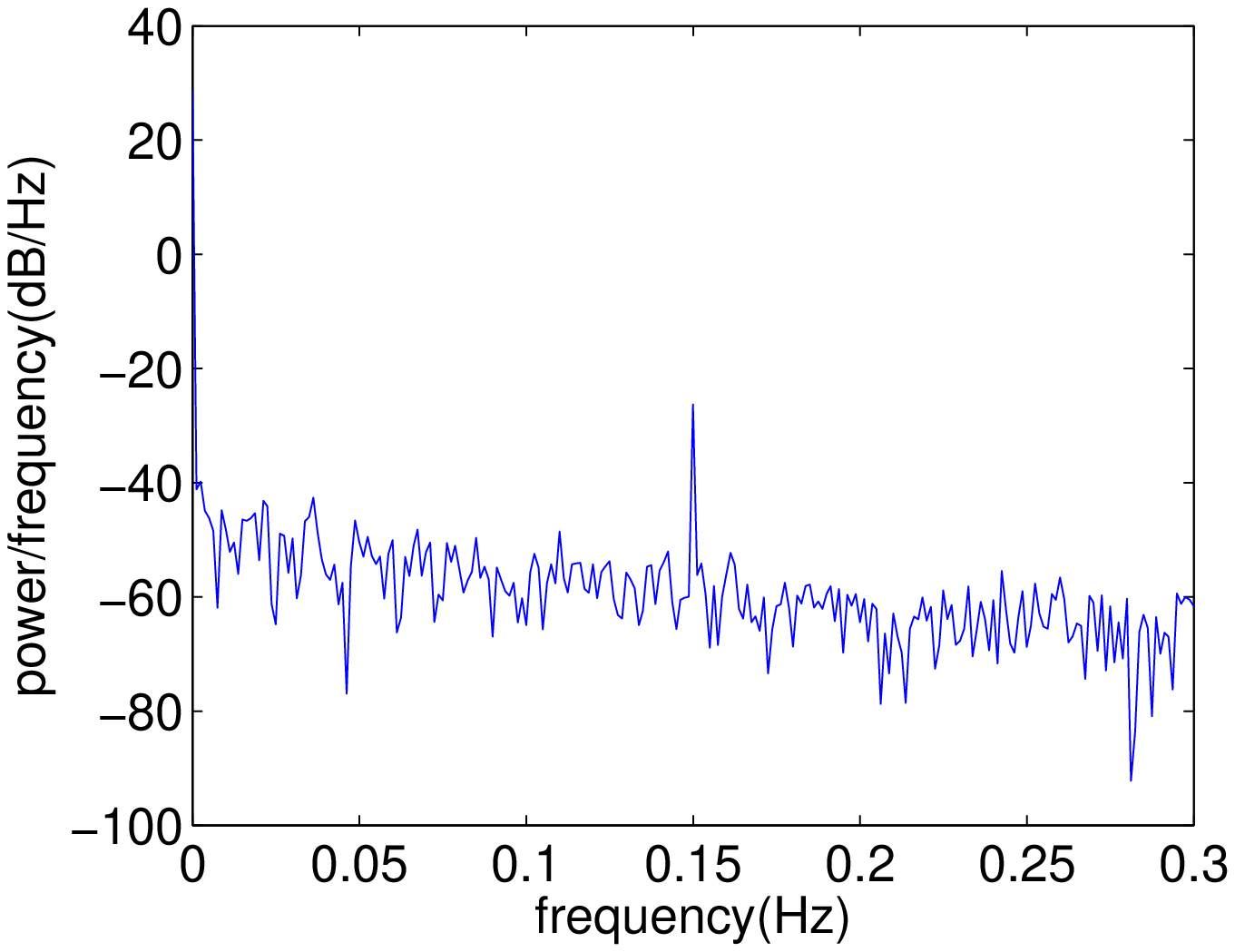}
\caption{}\label{ACPSDV5forced}
\end{subfigure}
\caption{(a). The moving kurtosis of the voltage magnitude shown in Fig. \ref{voltage-c}; (b). The power spectral density of the steady-state time series data between 700s and 1500s in Fig. \ref{voltage-c}.}
\end{figure}

In the last case, it is again hard to distinguish the oscillation mechanism directly from the time series data shown in Fig. \ref{voltage-c}. Likewise, following the proposed method, we firstly estimate the kurtosis to discern the weakly damped oscillation from the forced oscillation and the limit cycle. The moving kurtosis of the time series data with a window size of 50s is presented in Fig. \ref{kurtosisV5forced}, from which we can observe a significant change of the kurtosis before and after the variation.
Particularly, the kurtosis of the new steady-state time series data (between 700s and 1500s) is -0.74, 
and hence the weakly damped oscillation has been ruled out from all the mechanisms. We next estimate the power spectral density of the steady-state time series data between 700s and 1500s in Fig. \ref{voltage-c}. As presented in Fig. \ref{ACPSDV5forced}, there is a thin spike around 0.15Hz in the power spectrum. Hence, we conclude that the system is experiencing forced oscillation at a frequency around 0.15Hz.

\begin{figure}[!ht]
\centering
\begin{subfigure}[t]{0.5\linewidth}
\includegraphics[width=1.8in ,keepaspectratio=true,angle=0]{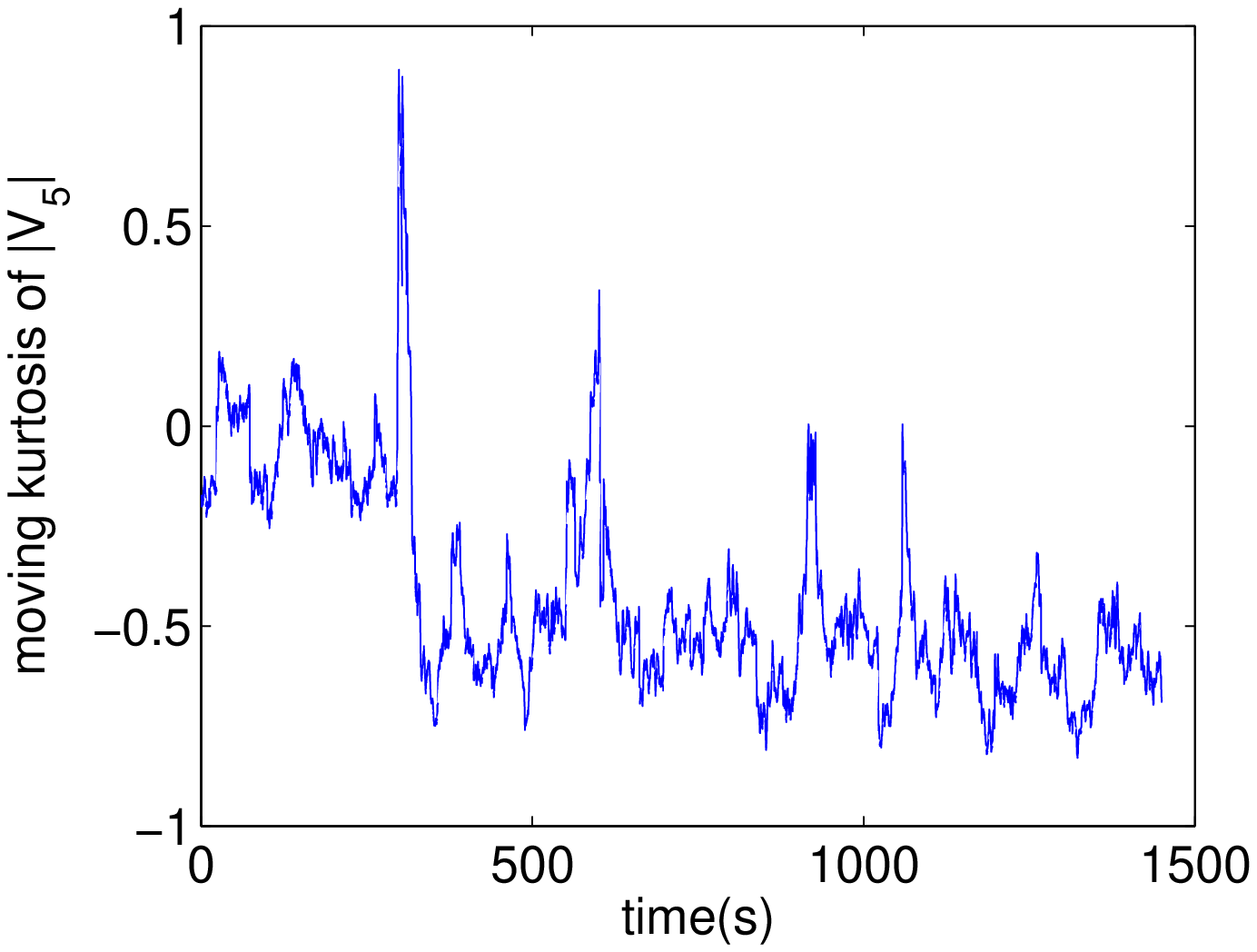}
\caption{}\label{kurtosisforced-noise}
\end{subfigure}%
\begin{subfigure}[t]{0.5\linewidth}
\includegraphics[width=1.8in ,keepaspectratio=true,angle=0]{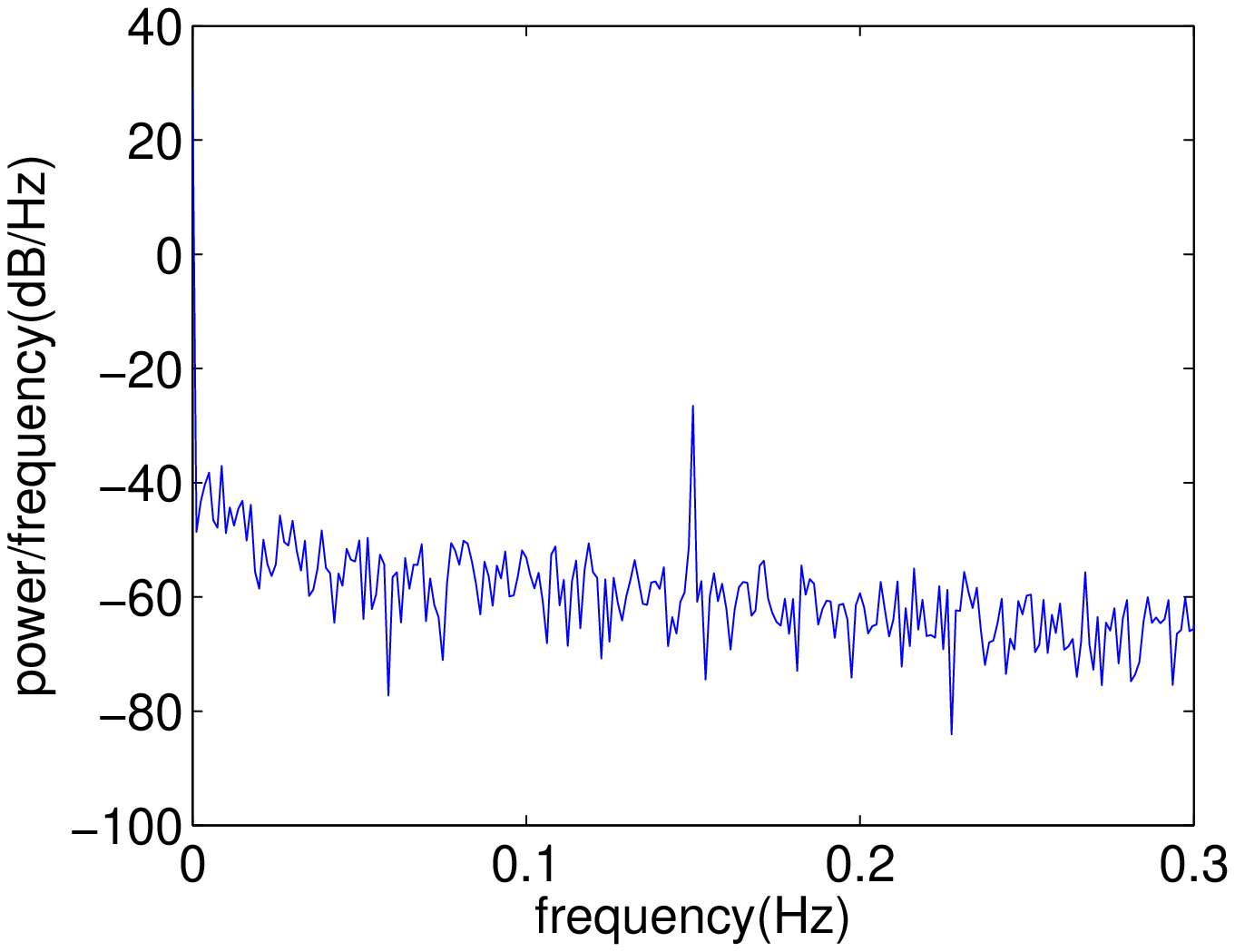}
\caption{}\label{ACPSDforced-noise}
\end{subfigure}
\caption{(a). The moving kurtosis of the voltage magnitude with measurement noise; (b). The power spectral density of the steady-state time series data between 700s and 1500s with measurement noise.}
\end{figure}

We also include measurement noise in simulations. A white Gaussian noise with $\mbox{STD}=10^{-3}$ is added to the simulated PMU data, and the corresponding moving kurtosis and PSD are shown in Fig. \ref{kurtosisforced-noise}-\ref{ACPSDforced-noise}. It is found that both the kurtosis and the PSD perform similarly as the case without measurement noise. Particularly, the moving kurtosis of the new steady-state time series data between 700s and 1500s is -0.56, and the power spectrum has a clear peak around 0.15Hz with a thin spike. Besides, the amplitude of the forced oscillation in the subspace of $|V_5|$ is $2\times 10^{-4}$ p.u., and variance of measurement noise is $10^{-6}$,  hence the SNR is approximately $-16.99$ dB, which is very low and will be even lower if the influence of load fluctuations is considered.

The actual situation is as follows, the exponential recovery load at Bus 5 increases by 5\% between 300s and 350s, then one cyclic load joins at Bus 5 with a forced frequency 0.15Hz. Therefore, the increasing fluctuation of the bus voltage shown in Fig. \ref{voltage-c} is due to forced oscillation.

This example shows that the proposed diagnosis method can successfully identify the forced oscillation among all the mechanisms at a low SNR. As mentioned before, The mechanism diagnosis for the oscillation with small amplitude is necessary to implement mitigation control in a timely manner.

\begin{table}[!ht]
\centering
\normalsize{
\begin{tabular}{|c|c|c|c|}
\hline
Bus 1&-0.33& Bus 8&-0.64\\
\hline
Bus 2&-0.48& Bus 9&-0.12\\
\hline
Bus 3&-0.29& Bus 10&-0.14\\
\hline
Bus 4&-0.60& Bus 11&-0.22\\
\hline
Bus 5&\textbf{-0.74}& Bus 12&-0.30\\
\hline
Bus 6&-0.40& Bus 13&-0.27\\
\hline
Bus 7&-0.60& Bus 14&-0.12\\
\hline
\end{tabular}
}
\caption{The kurtosis of voltage magnitudes at different buses for the new steady state (between 700s and 1500s) in \textit{Case III}.}\label{TablekurtosisForced}
\end{table}

Similarly, the kurtosis may help locate the oscillation source. The absolute value of the kurtosis indicates the amplitude of oscillation as shown in (\ref{forcedkurtosis}). By comparing the kurtosis of the voltage magnitudes at different buses, the location or region of the forced oscillation may be identified. In this example, the kurtosis of each voltage magnitude is shown in Table \ref{TablekurtosisForced}. For illustration purpose, we do not consider measurement noise here. We see that the kurtosis of the voltage magnitude at Bus 5 has the maximum absolute value, which indicates that the amplitude of voltage oscillation is larger at Bus 5 compared with other buses. It turns out that the cyclic load indeed is at Bus 5 with a forced frequency at 0.15Hz. Removing of the cyclic load at Bus 5 after the locating will then effectively eliminate the sustained oscillation. Further studies may be needed to design a more comprehensive method for locating the oscillation sources.

\section{validation on different cases}
In order to show that the proposed oscillation diagnostic method is robust, more simulation results are to be presented in this section. We change some parameters of the base case, and let the resulting system still experience Hopf bifurcation as the loads increase. Likewise, we also add a cyclic load to the base case in order to simulate forced oscillation. In particular, we set the active power of the exponential recovery load at Bus 13 to be 0.605 p.u. instead of 0.55 p.u. which is the value of the base case for \textit{Case I-III}. In other words, we make the active power of the exponential recovery load increase by 10\%.

\begin{figure}[!ht]
\centering
\begin{subfigure}[t]{0.5\linewidth}
\includegraphics[width=1.8in ,keepaspectratio=true,angle=0]{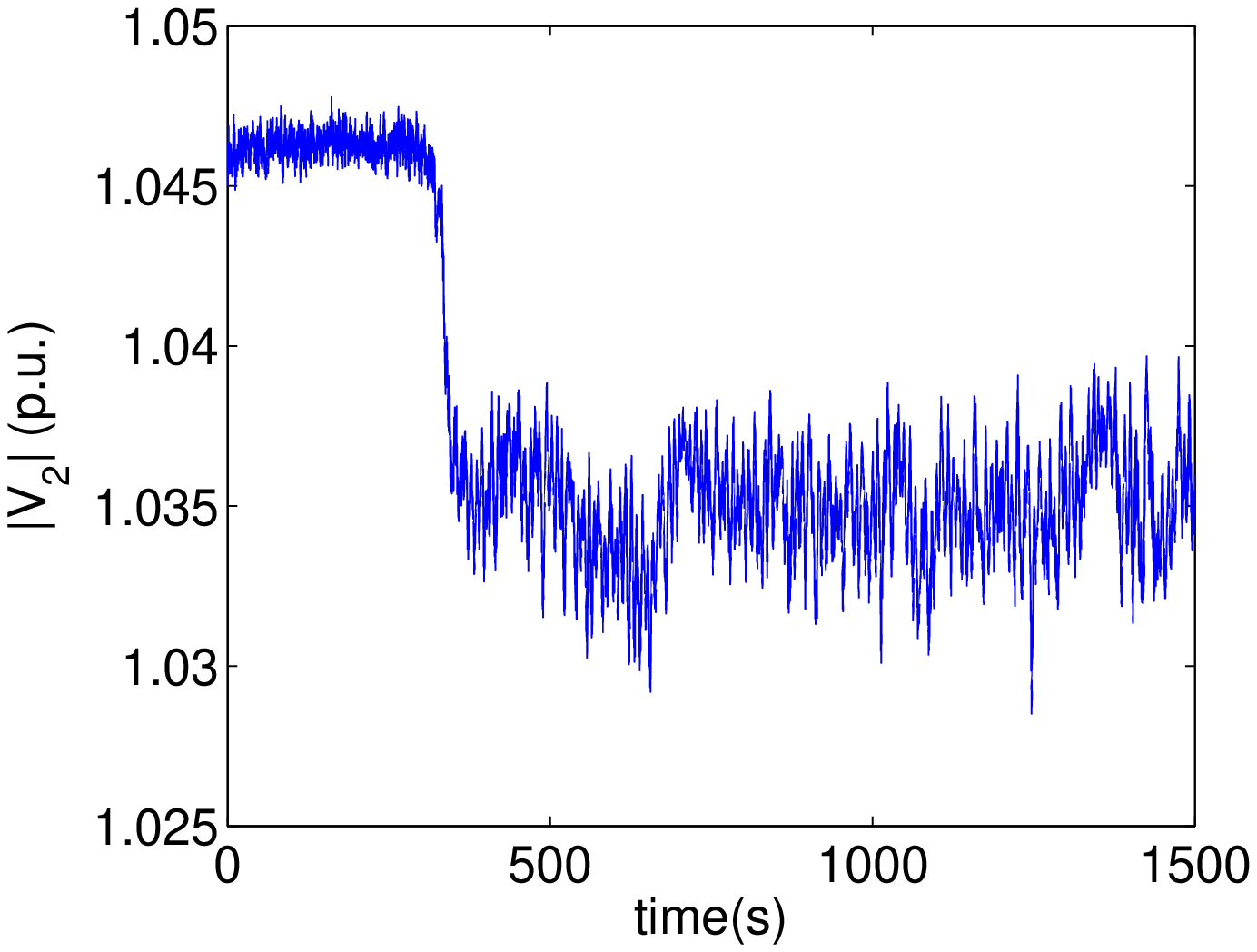}
\caption{Voltage magnitude at Bus 2}\label{voltage-a-1}
\end{subfigure}%
\begin{subfigure}[t]{0.5\linewidth}
\includegraphics[width=1.8in ,keepaspectratio=true,angle=0]{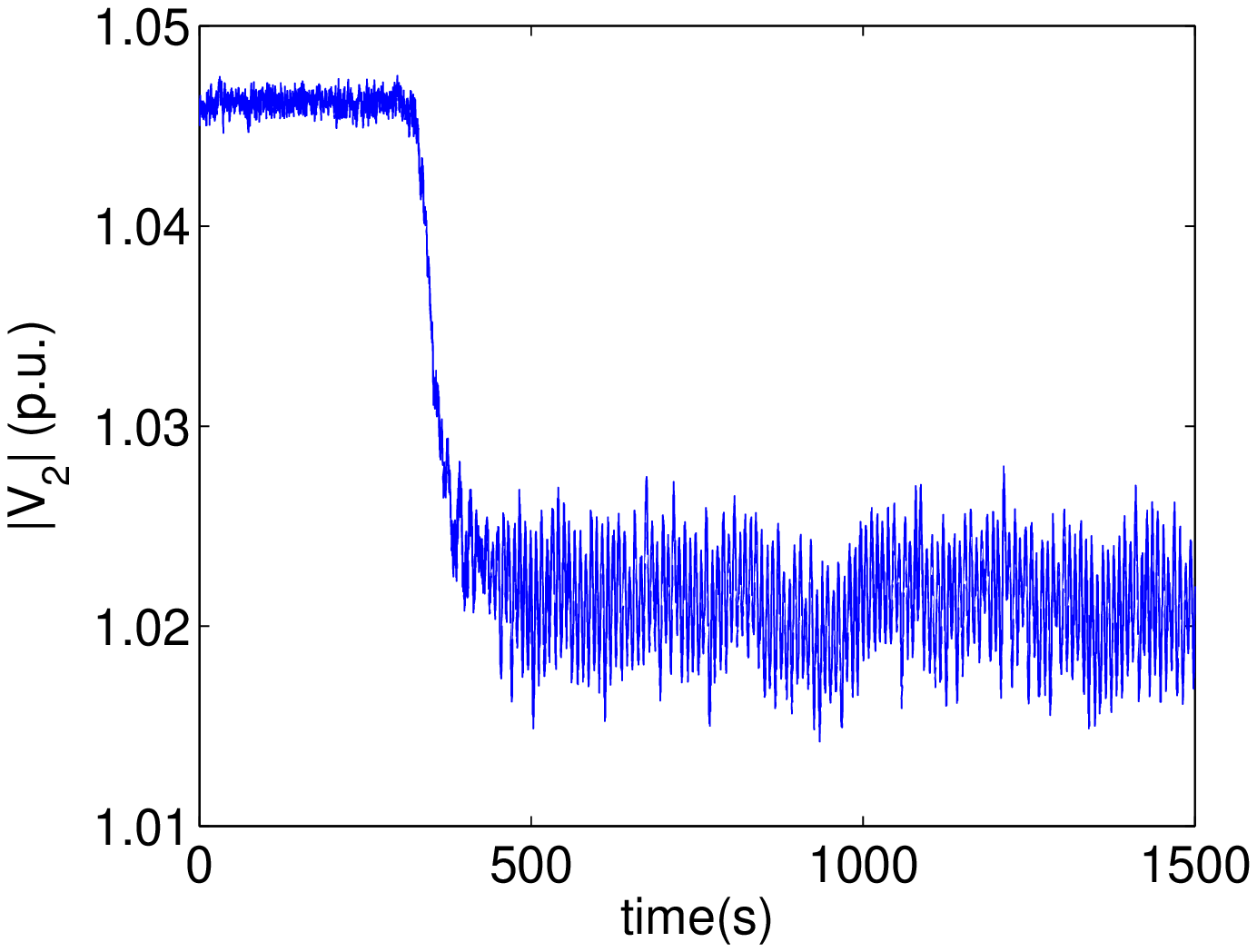}
\caption{Voltage magnitude at Bus 2}\label{voltage-b-1}
\end{subfigure}
\begin{subfigure}[t]{0.5\linewidth}
\includegraphics[width=1.8in ,keepaspectratio=true,angle=0]{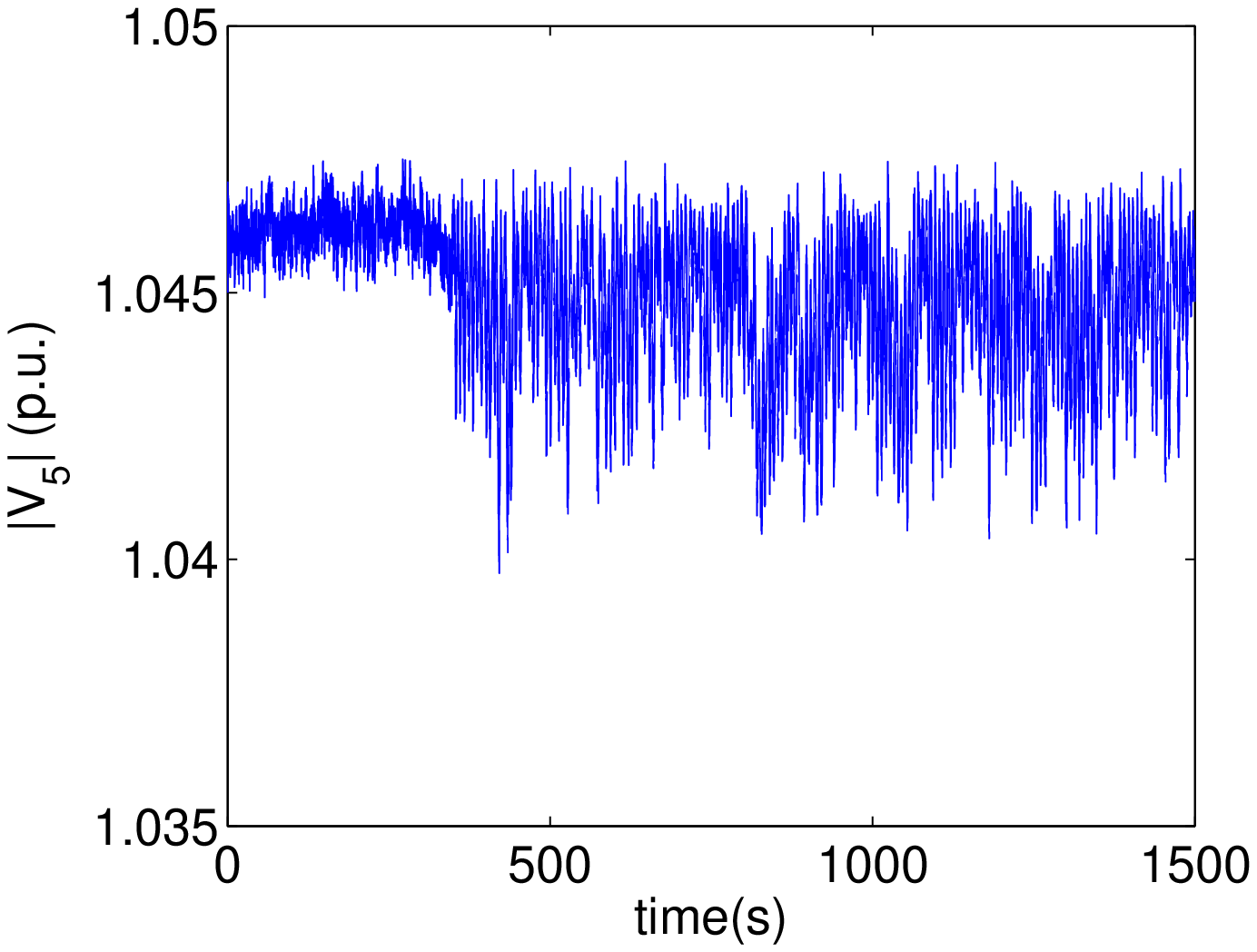}
\caption{Voltage magnitude at Bus 5}\label{voltage-c-1}
\end{subfigure}
\caption{Voltage magnitudes in different mechanisms.}\label{voltage-1}
\end{figure}

Starting from 300s, the exponential recovery loads at Bus 5 and Bus 12 increase at twice the speed of \textit{Case I}, and by 335s, both loads grow by 7\% and stop increasing afterwards. Because the load increasing stop before the Hopf bifurcation, the system has weakly damped oscillation. Particularly, Fig. \ref{voltage-a-1} shows the voltage magnitude of Bus 2 of this case. Furthermore,
if the increasing of exponential recovery loads at Bus 5 and Bus 12 do not stop until 360s, both loads grow by 12\%. By then, the system has passed the Hopf bifurcating point, and the voltage magnitude at Bus 2 is shown in Fig. \ref{voltage-b-1}. Regarding the forced oscillation, the manner of load changing is exactly the same as that in \textit{Case III}, and Fig. \ref{voltage-c-1} shows the voltage magnitude at Bus 5.

\begin{figure}[!ht]
\centering
\begin{subfigure}[t]{0.5\linewidth}
\includegraphics[width=1.8in ,keepaspectratio=true,angle=0]{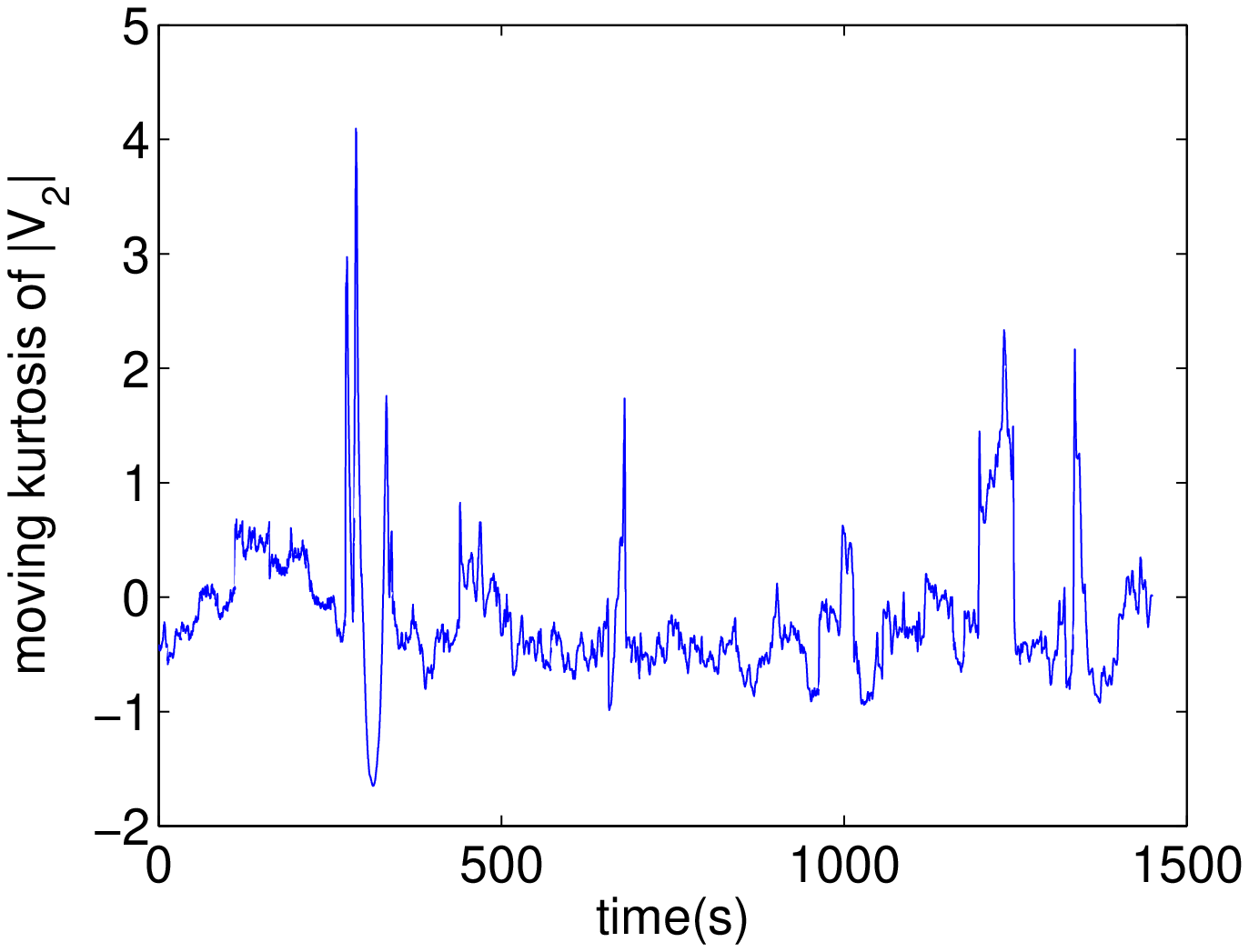}
\caption{}\label{kurtosisV2WD-1}
\end{subfigure}%
\begin{subfigure}[t]{0.5\linewidth}
\includegraphics[width=1.8in ,keepaspectratio=true,angle=0]{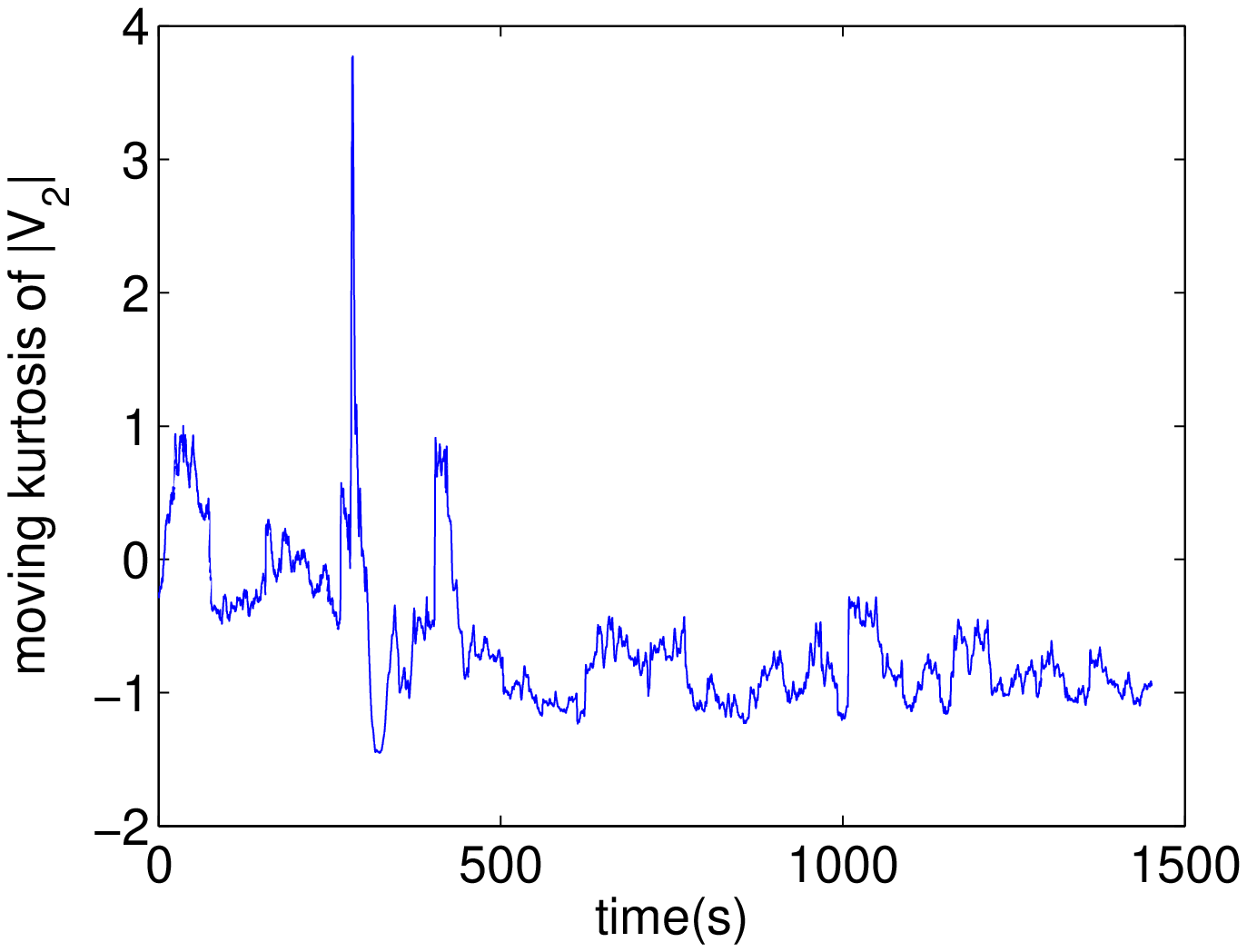}
\caption{}\label{kurtosisV2LC-1}
\end{subfigure}
\begin{subfigure}[t]{0.5\linewidth}
\includegraphics[width=1.8in ,keepaspectratio=true,angle=0]{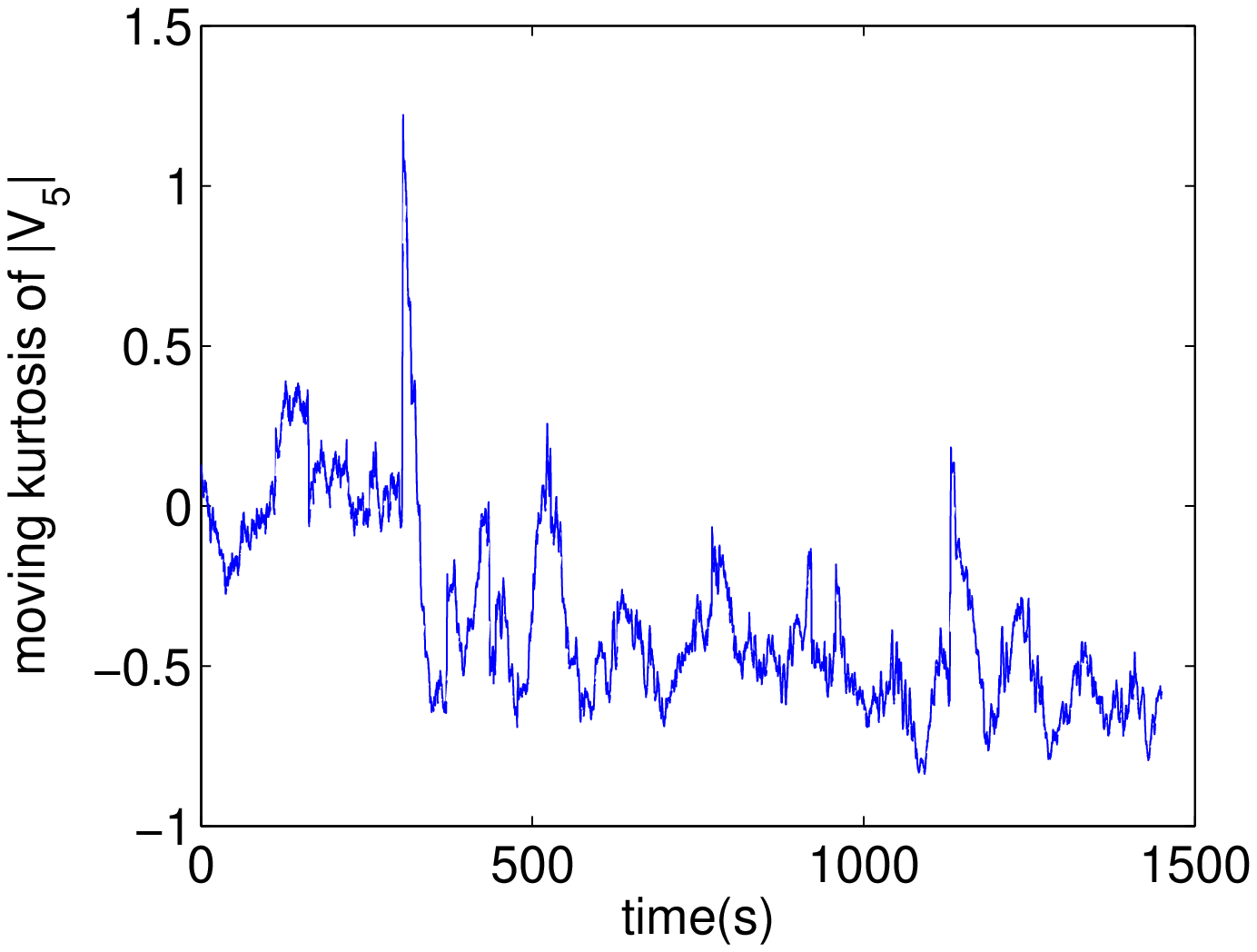}
\caption{}\label{kurtosisV5forced-1}
\end{subfigure}
\caption{(a). The moving kurtosis of the voltage magnitude shown in Fig. \ref{voltage-a-1}; (b). The moving kurtosis of the voltage magnitude in Fig. \ref{voltage-b-1}; (c). The moving kurtosis of the voltage magnitude shown in Fig. \ref{voltage-c-1};}
\end{figure}

We then follow the proposed oscillation diagnostic method and see whether the kurtosis and the PSD exhibit the expected characteristics. The moving kurtosis for the case before Hopf bifurcation is shown in Fig. \ref{kurtosisV2WD-1}, from which it is observed that the kurtosis is still close to zero after the transient period. Actually, the kurtosis of the new steady state (between 700s to 1500s) is -0.24. As for the case that has passed the Hopf bifurcating point, Fig. \ref{kurtosisV2LC-1} shows its moving kurtosis, from which it is seen that the kurtosis of the new steady state has been away from zero. In fact, the kurtosis of new steady state is -0.86. In the case of forced oscillation, the moving kurtosis is presented in Fig. \ref{kurtosisV5forced-1}. It is seen that the kurtosis is away from zero, which is actually -0.51 for the new steady state.
From these results, it is seen that kurtosis is a robust statistic to discern weakly damped oscillation from the other mechanisms.

\begin{figure}[!ht]
\centering
\begin{subfigure}[t]{0.5\linewidth}
\includegraphics[width=1.8in ,keepaspectratio=true,angle=0]{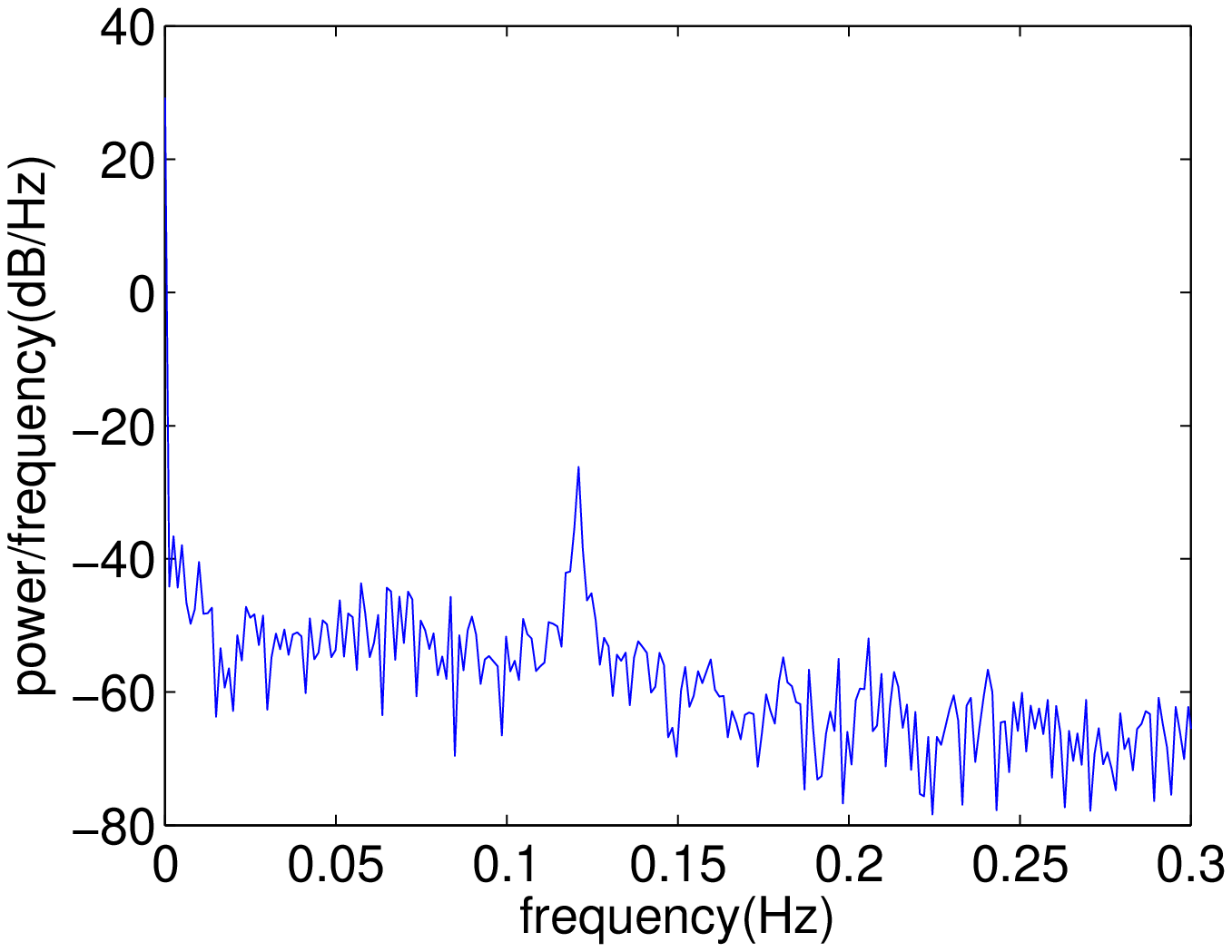}
\caption{}\label{ACV2LC-1}
\end{subfigure}%
\begin{subfigure}[t]{0.5\linewidth}
\includegraphics[width=1.8in ,keepaspectratio=true,angle=0]{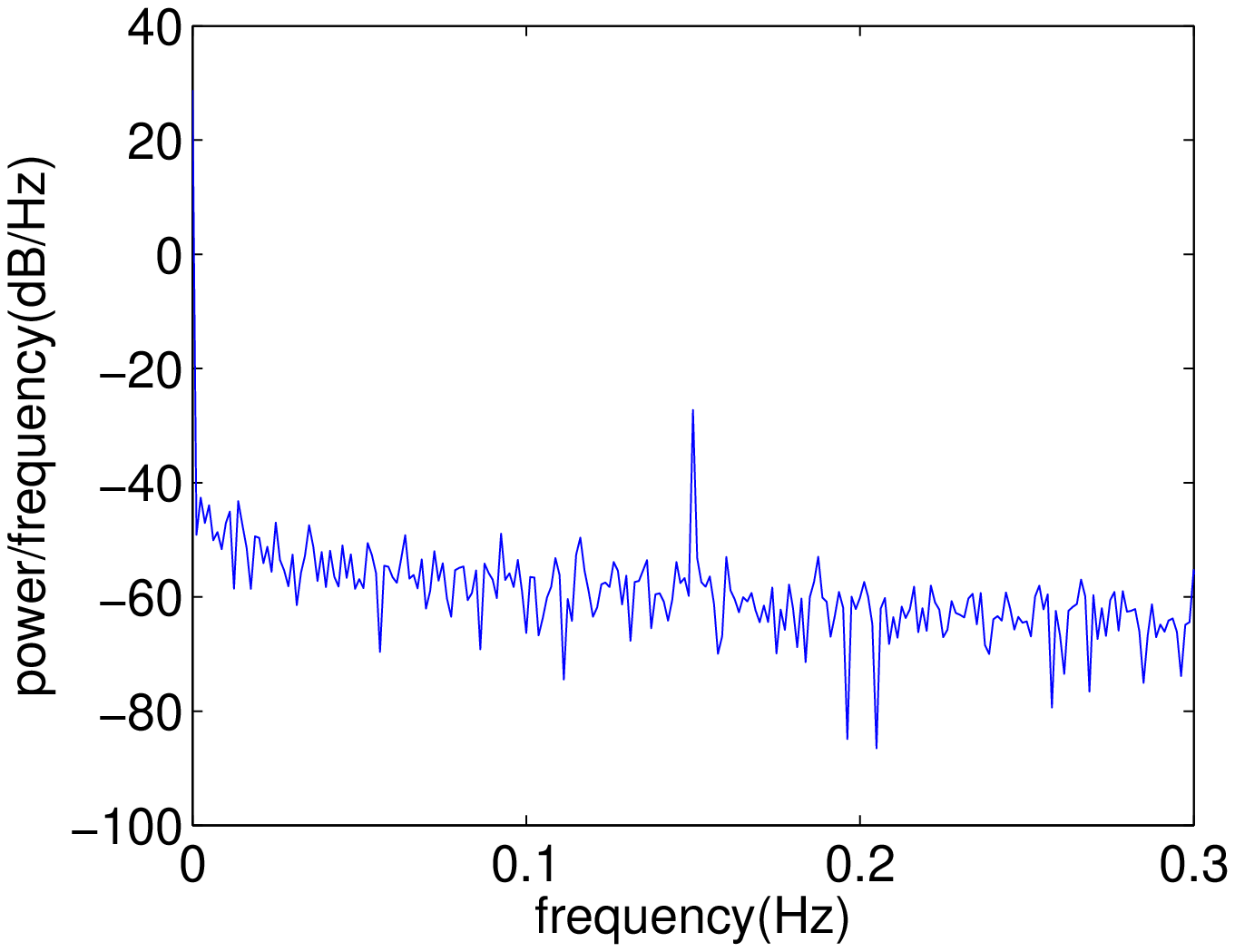}
\caption{}\label{ACV5forced-1-1}
\end{subfigure}
\caption{(a). The power spectral density of the new steady state of the time series data in Fig. \ref{voltage-b-1}. (b). The power spectral density of the new steady state of the time series data in Fig. \ref{voltage-c-1}.}
\end{figure}

We further estimate the PSD for the cases of limit cycle and forced oscillation. The results are shown in Fig. \ref{ACV2LC-1}-\ref{ACV5forced-1-1}. It is observed that the PSD of the limit cycle has a wide bandwidth, whereas the PSD of the forced oscillation exhibits a thin spike around 0.15HZ. Those results further validate that the PSD is able to distinguish the forced oscillation from the limit cycle.

The extra simulation results given in this section are to demonstrate that the proposed oscillation diagnostic method is robust to the change of system parameters.

\section{Conclusions and Perspectives}\label{sectionconclusion}
This paper elaborates the mechanisms of power system oscillations in a unified mathematical framework, under which the statistical signatures of different oscillation mechanisms are investigated. Even though oscillations with different mechanisms look alike in time series data, they exhibit distinct statistical signatures based on which an oscillation diagnosis method is developed. Particularly, the oscillation diagnosis method utilize the kurtosis to discern the weakly damped oscillation from the others, and then use the power spectral density to differentiate the limit cycle and the forced oscillation. Numerical example show that the proposed method can accurately identify the exact mechanism of sustained oscillation using PMU data, and also provide insights towards locating the oscillation sources. Though control actions are briefly discussed for each mechanism, further investigations are needed to better locate the problematic components and design the right control actions after identifying the exact oscillation mechanism.

\appendices
\section{Derivation of Kurtosis}\label{kurtosisderivation}
The kurtosis of $x(t)=(\sqrt{\gamma}+p(t))\cos(\phi(t))$ can be derived as follows. Denote $\overline{{x}(t)}=\frac{1}{T}\int_{t}^{t+T}x(t)dt$, then we have:
\begin{equation}
\E[{x}(t)]=\overline{(\sqrt{\gamma}+p(t))\cos(\phi(t))}=0\\
\end{equation}
\begin{equation}
\Var[{x}(t)]=\overline{((\sqrt{\gamma}+p(t))\cos(\phi(t)))^2}=\frac{1}{2}(\gamma+\overline{p^2(t)})\\
\end{equation}
\begin{eqnarray}
\E[(x(t)^4]&=&\overline{(\gamma^2+6\gamma\overline{p^2(t)}+\overline{p^4(t)})\cos^4(\phi(t))}\nonumber\\
&=&\frac{3}{8}(\gamma^2+6\gamma\overline{p^2(t)}+3(\overline{p^2(t)})^2)\label{4moment}
\end{eqnarray}
The derivation of (\ref{4moment}) uses the fact that $x_1^4(t)=3(\overline{x_1^2(t)})^2$ for Gaussian process and odd moments of Gaussian process are all zero.
The kurtosis of $x(t)$ is:
\begin{equation}
\mathrm{Kurt}[x(t)]=-\frac{3[(\gamma-\overline{p^2(t)})^2-2(\overline{p^2(t)})^2)]}{2(\gamma+\overline{p^2(t)})^2}
\end{equation}

Following similar procedures for the limit cycle, the kurtosis of $x(t)=\rho\cos(\Omega t)+x_1(t)$ for the forced oscillation can be derived as follows.
\begin{equation}
\E[{x}(t)]=\overline{\rho\cos(\Omega t)+x_1(t)}=0\\
\end{equation}
\begin{equation}
\Var[{x}(t)]=\overline{(\rho\cos(\Omega t)+x_1(t))^2}=\frac{\rho^2}{2}+\overline{x_1^2(t)}\\
\end{equation}
\begin{eqnarray}
\E[(x(t)^4]&=&\overline{\rho^4\sin^4(\Omega t)+6\rho^2\sin^2(\Omega t)x_1^2(t)+x_1^4(t)}\nonumber\\
&=&\frac{3}{8}\rho^4+3 \rho^2\overline{x_1^2(t)}+3(\overline{x_1^2(t)})^2
\end{eqnarray}
Therefore, the kurtosis of $x(t)$ is:
\begin{equation}
\mathrm{Kurt}[x(t)]=\frac{\E[(x(t)-\rho)^4]}{\Var^2[x(t)]}-3=-\frac{3}{2}\frac{1}{(1+2\frac{\overline{x_1^2(t)}}{\rho^2})^2}
\end{equation}

\section{parameter values of the base case}\label{appendixbasecase}
The base case is modified based on the test system ``d{\_}014{\_}dyn\_mdl'' in PSAT-2.1.8 \cite{Milano:article}. The line between Bus 7-9 has been deleted. The active power of the load at Bus 2 has been increased to 0.55 p.u. from 0.189 p.u., and the active power of the load at Bus 9 has been increase to 0.6 p.u. from 0.413 p.u.. All loads are modelled as exponential recovery loads whose parameter values are given in Table \ref{apendixtable1}. Besides, there are two turbine governors for the generators at Bus 1 and 2, with parameter values shown in Table \ref{appedixtable2}. All generators are also controlled by OXLs whose parameter values are given in Table \ref{appedixtable3}.

\begin{table}[!ht]
\centering
\caption{Exponential Recovery Load Parameter Values}{\label{apendixtable1}}
\begin{tabular}{|c|c|}
\hline
Parameter&Value\\
\hline
active power percentage $k_p$&$100\%$\\
\hline
reactive power percentage $k_q$&$100\%$\\
\hline
active power time constant $T_p$&$10$s\\
\hline
reactive power time constant $T_q$&$10$s\\
\hline
static active power exponent $\alpha_s$&$1$\\
\hline
dynamic active power exponent $\alpha_t$&$1.5$ for the load at Bus 4,5,9,11\\
&$5$ for the other loads \\
\hline
static reactive power exponent $\beta_s$&$2$\\
\hline
dynamic reactive power exponent $\beta_t$&$2.5$ for the load at Bus 4,5,9,11\\
&$10$ for the other loads\\
\hline
\end{tabular}
\end{table}

\begin{table}[!ht]
\centering
\caption{Turbine Governor Parameter Values}\label{appedixtable2}
\begin{tabular}{|c|c|}
\hline
Parameter&Value\\
\hline
reference speed $\omega^{0}_{ref}$&$1$ p.u.\\
\hline
droop $R$&$0.02$ p.u.\\
\hline
maximum turbine output $p^{max}$&$1.2$ p.u.\\
\hline
minimum turbine output $p^{min}$&$0.3$ p.u.\\
\hline
governor time constant $T_s$&$0.1$s\\
\hline
servo time constant $T_c$&$0.45$s\\
\hline
transient gain time constant $T_3$&$0$s\\
\hline
power fraction time constant $T_4$&$12$s\\
\hline
reheat time constant $T_5$&$50$s\\
\hline
\end{tabular}
\end{table}

\begin{table}[!ht]
\centering
\caption{Over Excitation Limiter Parameter Values}\label{appedixtable3}
\begin{tabular}{|c|c|}
\hline
Parameter&Value\\
\hline
maximum field current $i_f^{lim}$&$5.1$ p.u., $3.8$ p.u., $2.5$ p.u., \\
for OXL 1-5&$3.2$ p.u., $5.8$ p.u.\\
\hline
integrator time constant $T_0$&$30$s for Generator 3\\
&$10$s for the others\\
\hline
maximum output signal $v_{\mbox{oxl}}$&$100$ p.u.\\
\hline
\end{tabular}
\end{table}
\section*{Acknowledgment}
The authors would like to thank Dr. Slava Maslennikov, Prof. Janusz Bialek and Prof. Paul Hines for helpful comments and discussions.

%
%

\end{document}